\documentclass[12pt]{article}
\usepackage{amsmath,latexsym,exscale,amssymb}
\newcommand {\al}   {\alpha}       \newcommand {\bt}  {\beta}
\newcommand {\g }   {\gamma}       \newcommand {\G }  {\Gamma}
\newcommand {\dl}   {\delta}       \newcommand {\e }  {\epsilon}
\newcommand {\z }   {\zeta}        \newcommand {\et}  {\eta}
\newcommand {\ve}   {\varepsilon}  
\newcommand {\lm}   {\lambda}      
          
\newcommand {\s }   {\sigma}      
         
\newcommand {\f }   {\varphi}      
         \newcommand {\om}  {\omega}
\newcommand {\Lm}   {\Lambda}      \newcommand {\Om}  {\Omega}
       
\newcommand {\pl}   {\partial}     \newcommand {\nb}  {\nabla}
\newcommand   {\const}{{\sf const}}   
     
\renewcommand {\det}{{\sf det}}       \renewcommand {\exp}{{\sf \,exp\,}}

\renewcommand {\ln}{{\sf ln\,}}         
\renewcommand {\sin}{{\sf sin}}       \renewcommand {\cos}{{\sf cos}}

\newcommand   {\arctg}{{\sf arctg}}
\newcommand   {\sh}{{\sf sh\,}}         \newcommand   {\ch}{{\sf ch\,}}

\newcommand {\CU }  {{\cal U}}      
\newtheorem{Theorem}{Theorem}
\begin{document}
\title     {Effective action for scalar fields\\
            in two-dimensional gravity}
\author    {M. O. Katanaev
            \thanks{E-mail: katanaev@mi.ras.ru}\\ \\
            \sl Steklov Mathematical Institute,\\
            \sl Gubkin St. 8, Moscow 117966, Russia}
\date      {1 October 2001}
\maketitle
\begin{abstract}
  We consider a general two-dimensional gravity model minimally or
  nonminimally coupled to a scalar field. The canonical form of the
  model is elucidated, and a general solution of the equations of
  motion in the massless case is reviewed. In the presence of a
  scalar field all geometric fields (zweibein and Lorentz connection)
  are excluded from the model by solving exactly their Hamiltonian
  equations of motion. In this way the effective equations of motion
  and the corresponding effective action for a scalar field are obtained.
  It is written in a Minkowskian space-time and does not include any
  geometric variables. The effective action arises as a boundary term
  and is nontrivial both for open and closed universes. The reason is
  that unphysical degrees of freedom cannot be compactly supported
  because they must satisfy the constraint equation. As an example we
  consider spherically reduced gravity minimally coupled to a massless
  scalar field. The effective action is used to reproduce the Fisher
  and Roberts solutions.
\end{abstract}
\section{Introduction                                 \label{sintro}}
In recent years great attention has been paid to two-dimensional gravity
models mainly for two reasons: a close relation to string theory and good
laboratories to get deeper insights in classical and quantum properties
of gravity models. The first and the simplest two-dimensional constant
curvature gravity was proposed in \cite{BaNeCh79}. It attracted much
interest after the papers \cite{Jackiw84,Teitel83}. Constant curvature
surfaces are described by the integrable Liouville equation, and the main
concern was given to physical interpretation of the solutions and
inclusion of matter fields.

In the present paper we consider a wide class of two-dimensional gravity
models with torsion and equivalent generalized dilaton models.
A two-dimensional gravity model with torsion was proposed in
\cite{KatVol86,VolKat86B,KatVol90} to provide dynamics for the
metric on a string world sheet already at the classical level.
Two-dimensional dilaton gravity \cite{Witten91,MaSeWa91} was proposed
as an effective model coming from the string theory. Both models turned
out to be integrable. A general solution to the equations of motion of
two-dimensional gravity with torsion was first given in the conformal
gauge \cite{Katana89B,Katana90,Katana91}. All solutions are divided into
two classes: (i) constant curvature and zero torsion and (ii) nonconstant
curvature and nonzero torsion. In this way constant curvature gravity
models are included in two-dimensional gravity with torsion with a well
defined purely geometric action and a natural way of introduction of
matter fields. The equations of motions of two-dimensional gravity with
torsion were integrated in the light-cone gauge \cite{KumSch92A} and
without any gauge fixing \cite{Solodu93A,MiGrObTrHe93,SchStr94A}.
At that time the equivalence of two-dimensional gravity with torsion
and dilaton gravity was unknown, and solution of the equations of
motion for a generalized dilaton gravity was independently obtained
in the light-cone gauge \cite{Filipp96}.

Later it was realized that in fact the two-dimensional gravity models with
torsion and dilaton gravity models are equivalent
\cite{KaKuLi96,KuLiVa97,KumTie98}.
From a geometric point of view the dilaton field is the momentum canonically
conjugate to the space component of the Lorentz connection. The equivalence
of these models is nontrivial because they contain a different set of fields.

Throughout this paper we consider a general two-dimensional gravity model
quadratic in torsion.
The integrability of the model is connected to the existence of a
Killing vector field in the absence of matter. It is worth noting that
without matter fields the two-dimensional gravity models do not describe
any propagating degree of freedom. This suggests a generalization of the
models by addition of matter fields to attack the problems arising in
black hole formation and quantum gravity. Unfortunately, addition of
matter destroys integrability. Some exact solutions of two-dimensional
gravity with torsion coupled to scalar fields were found
\cite{Obukho94,ObSoMi94,PelStr98}.
We mention also nontrivial solutions for dilaton gravity coupled
to scalar fields found in \cite{Filipp96,FilIva98,Cavagl98}.
In Section \ref{sgenla} we consider a Lagrangian
with arbitrary dependence on scalar curvature and torsion. There we clarify
the statement that this general model yields integrable equations of motion
made in the literature \cite{MiGrObTrHe93,Solodu94A,KloStr96A}.

Let us note that the two-dimensional gravity model nonminimally coupled
to a scalar field is important for general relativists working in four
dimensions. It is well known that spherically reduced general relativity
minimally coupled to a scalar field is equivalent to the dilaton gravity
model nonminimally coupled to a scalar field in two-dimensional space-time
\cite{ThIsHa84}. This model may be the simplest one to describe the
dynamics of a black hole formation by spherical scalar waves. Recently
it attracted much interest due to the discovery of critical phenomena
in the black hole formation \cite{Choptu93} (for review see \cite{Gundla00})
found in numerical simulations. A satisfactory analytical calculation of the
critical exponent and scaling period of discrete self-similarity has been
missing up to now, and hence our understanding of critical phenomena should
be considered incomplete.

In the present paper we consider the two-dimensional gravity model with
torsion or the equivalent dilaton gravity coupled to a scalar field.
We admit nonminimal coupling to be sufficiently general to include
spherically reduced gravity. After
describing the Lagrangian we elucidate the Hamiltonian formulation of
the model in full detail. The canonical form is essential for our
approach and I am not aware how to write down the effective action in
the Lagrangian formulation. Afterwards a general solution of the
equations of motion in the matterless case and the equivalence with
the dilaton model are briefly reviewed. In the presence of a scalar field
the equations of motion are not integrable. Nevertheless they may be
partly integrated. We solve the geometric part of the equations of
motion with respect to the zweibein and Lorentz connection assuming
the scalar field to be arbitrary. Then the solution is substituted
into the equations of motion for a scalar field. In this way we obtain
the effective equations of motion only for a scalar field and its
conjugate momentum. Next we show that the effective action yielding
these equations arises as a boundary term. It appears because unphysical
variables cannot be compactly supported functions as the consequence
of constraint equations. The effective equations of motion are written
in a Minkowskian space-time and provide a general solution to the whole
problem because the metric can be easily reconstructed for a given
solution to the effective equations of motion. The effective action
for spherically reduced gravity in a special case was first derived in
\cite{Unruh76}. Here we generalize this action by considering the more
general two-dimensional gravity part and arbitrary coupling to scalars.
For spherically reduced gravity we give a more general effective action
depending on one arbitrary function on time to be fixed by boundary
conditions on the metric.

The reduction of the whole system of the equations of motion to the
equations for a scalar field and its conjugate momentum does not provide
a general solution of the model because the effective equations are
integro-differential and complicated. As an example I considered
spherically reduced gravity. The effective action has clear limiting
cases and for a small scalar field reduces to an ordinary action for
spherical waves on a Minkowskian or Schwarzschild background. In the
static case the equations may be integrated in elementary functions
and yield the same solution as found by Fisher \cite{Fisher48}.
As an example we consider also the Roberts solution \cite{Robert89}
which provides a self-similar solution to the effective equations of
motion.
\section{The action                                    \label{smolag}}
Let us consider a Lorentzian surface with coordinates $x^\al=\{\tau,\s\}$,
$\al=0,1$. We assume that it is equipped with a Riemann--Cartan geometry
defined by a zweibein $e_\al{}^a(x)$ and a Lorentz connection $\om_\al(x)$.
A general type action for a scalar field coupled to two-dimensional
gravity has the form
\begin{equation}                                        \label{efoact}
  S=\int\! dx^2 (L_G+L_X),
\end{equation}
where the Lagrangian for a scalar field $X(x)$ of mass $m=\const$ is
\begin{equation}                                        \label{elagst}
  L_X=-\frac12\rho e(\pl X^2-m^2X^2),
\end{equation}
and we introduce shorthand notations
$$
  \pl X^2=g^{\al\bt}\pl_\al X\pl_\bt X,~~~~~~e=\det e_\al{}^a.
$$
For
$\rho=\const$ we have a minimally coupled scalar. This Lagrangian is
easily generalized to a bosonic string moving in $D$-dimensional
space-time. To this end one has to replace a scalar by a set of scalars
$X^\mu(x)$, $\mu=0,1,\dots,D-1$ and set $m=0$.
The geometric part of the Lagrangian is given by a two-dimensional
gravity with torsion written in the first order form
\begin{equation}                                        \label{egrato}
  L_G=-\frac12(\pi\widehat R+p_a\widehat T^{*a})
      -e\left(\frac12p_ap^aU+V\right),
\end{equation}
where the densities of two-dimensional scalar curvature and the
pseudotrace of torsion are given by
\begin{align}                                           \label{escden}
  \widehat R&=eR=2\hat\ve^{\al\bt}\pl_\al\om_\bt,
\\                                                      \label{epttod}
  \widehat T^{*a}&=eT^{*a}
  =2\hat\ve^{\al\bt}(\pl_\al e_\bt{}^a-\om_\al\ve^a{}_b e_\bt{}^b).
\end{align}
Here $\hat\ve^{\al\bt}=e\ve^{\al\bt}=e\ve^{ab}e^\al{}_ae^\bt{}_b$ is the
totally antisymmetric tensor density, $\hat\ve^{01}=-\hat\ve^{10}=-1$,
$\pi$ and $p_a$ are considered as independent variables, and $U=U(\pi)$
and $V=V(\pi)$ are arbitrary functions of $\pi$. In the second order form,
when it exists, this Lagrangian is quadratic in torsion and arbitrarily
depends on the scalar curvature. A hat over a symbol means that it is a
tensor density of weight $-1$. For the nonminimally coupled scalar field
we have $\rho=\rho(\pi)$.

The case $U=0$ and $V=0$ describes surfaces of zero torsion and curvature
and is not interesting from the geometric point of view. Therefore we
assume that at least one of the functions differs from zero.
The Lagrangian (\ref{egrato}) is quite general. If $U=0$ then one has
the gravity model with zero torsion, and the Lagrangian in the second
order form is an arbitrary function of the scalar curvature defined by
$V$. For $U=1$ and $V=\pi^2+\const$ one immediately recovers
two-dimensional gravity with torsion quadratic in curvature and torsion
\cite{KatVol86,VolKat86B,KatVol90}. It is essentially a unique purely
geometric invariant model yielding second order equations of motion for
the zweibein and Lorentz connection. In this case $\pi$ and $p_a$ are
proportional to the scalar curvature and the pseudotrace of torsion
provided equations of motion are fulfilled.

For arbitrary functions $U(\pi)$ and $V(\pi)$ when matter fields do not
interact with the Lorentz connection $\om_\al$ the latter can be
excluded from the model by the use of its algebraic equations of motion
(Section \ref{sdilto}). This leads to a generalized two-dimensional dilaton
gravity, the function $\pi(x)$ being the dilaton field.
Then the matter Lagrangian (\ref{elagst}) describes scalars minimally,
$\rho=\const$, or nonminimally, $\rho=\rho(\pi)$, coupled to dilaton
gravity. Among these models of particular interest is the spherically
reduced gravity (see Section \ref{sredgr}) for which
\begin{equation}                                         \label{echfsr}
  U=\frac1{2\pi},~~~~V=-2\kappa K-\frac{\Lm\pi}\kappa,~~~~
  \rho=\frac\pi{2\kappa},
\end{equation}
where $K=1,0,-1$ for spherical, planar, or hyperbolic reductions,
respectively \cite{KaKlKu99}. $\Lm$ is the four-dimensional cosmological
constant, and $\kappa>0$ is the inverse gravitational constant in four
dimensions.

Note that in a general case the algebraic equation of motion for $\pi$ can
not be solved in elementary functions and the action in the second order
form cannot be written explicitly. An important point is that a gravity
with torsion without matter fields as given by (\ref{egrato}) can be
solved exactly for arbitrary functions $U$ and $V$ (see Section
\ref{sintmo}).
\section{Equations of motion                           \label{smoeqm}}
Equations of motion following from the action (\ref{efoact}) can be
written in the form
\begin{align}                                           \label{eqmopi}
  \frac1e\frac{\dl S}{\dl\pi}&:&-&\frac12R
  -\left(\frac12p^ap_a U'+ V'\right)-\frac12\rho'(\pl X^2-m^2X^2)=0,
\\                                                      \label{eqmotp}
  \frac1e\frac{\dl S}{\dl p_a}&:&-&\frac12T^{*a}-p^a U=0,
\\                                                      \label{eqmolc}
  \hat\ve_{\bt\al}\frac{\dl S}{\dl\om_\bt}&:&&
  \pl_\al\pi-p_a\ve^a{}_be_\al{}^b=0,
\\                                                      \label{eqmozw}
  \hat\ve_{\bt\al}\frac{\dl S}{\dl e_\bt{}^a}&:&&\nb_\al p_a
  +\ve_{\al a}\left(\frac12p^bp_bU+V\right)+\rho T^\bt{}_a\ve_{\bt\al}=0,
\\                                                      \label{eqmozx}
  \frac1e\frac{\dl S}{\dl X}&:&&\rho(
  g^{\al\bt}\widetilde\nb_\al\widetilde\nb_\bt X+m^2X)
  +\rho' g^{\al\bt}\pl_\al\pi\pl_\bt X=0,
\end{align}
where
$$
  \nb_\al p_a=\pl_\al p_a-\om_\al\ve_a{}^b p_b
$$
is the covariant derivative, and
\begin{equation}                                        \label{enemot}
  T_{\al\bt}=\pl_\al X\pl_\bt X-\frac12g_{\al\bt}(\pl X^2-m^2X^2)
\end{equation}
is the energy-momentum tensor of the scalar field. Primes on
functions $U,V,$ and $\rho$ always mean derivatives with respect to the
argument $\pi$. Transformation of Greek indices into Latin ones
and vice versa is everywhere
performed using the zweibein field and its inverse, and $\widetilde\nb$
denotes the covariant derivative with Christoffel's symbols.

The action (\ref{efoact}) is invariant under local Lorentz rotations
and general coordinate transformations which produce linear relations
between the equations. Local Lorentz rotations by the angle $\om(x)$
\begin{align}                                           \nonumber
  \dl e_\al{}^a&=-e_\al{}^b\ve_b{}^a\om, & \dl\om_\al&=\pl_\al\om,
\\
  \dl p_a&=\ve_a{}^b p_b\om, & \dl\pi&=0,
\\                                                      \nonumber
  \dl X&=0 &&
\end{align}
lead to the following dependence between equations of motion
\begin{equation}                                        \label{edeplo}
  \widetilde\nb_\al\frac{\dl S}{\dl\om_\al}
  +\frac{\dl S}{\dl e_\al{}^a}e_\al{}^b\ve_b{}^a
  -\frac{\dl S}{\dl p_a}\ve_a{}^bp_b=0.
\end{equation}
Variations of the fields under general coordinate
transformations with parameters $\e^\al(x)$
\begin{align}                                          \nonumber
  \dl e_\al{}^a&=-\pl_\al\e^\bt e_\bt{}^a-\e^\bt\pl_\bt e_\al{}^a, &
  \dl\om_\al&=-\pl_\al\e^\bt\om_\bt-\e^\bt\pl_\bt\om_\al,
\\
  \dl p_a&=-\e^\bt\pl_\bt p_a, & \dl\pi&=-\e^\bt\pl_\bt\pi,
\\                                                     \nonumber
  \dl X&=-\e^\bt\pl_\bt X,
\end{align}
together with the identity (\ref{edeplo}) produce two linear relations
between equations of motion
\begin{equation}                                        \label{edegco}
  e_\al{}^a\nb_\bt\frac{\dl S}{\dl e_\bt{}^a}
  -\frac{\dl S}{\dl e_\bt{}^a}T_{\al\bt}{}^a
  +\frac12\frac{\dl S}{\dl\om_\bt}\ve_{\al\bt}R
  -\frac{\dl S}{\dl\pi}\pl_\al\pi
  -\frac{\dl S}{\dl p_a}\nb_\al p_a-\frac{\dl S}{\dl X}\pl_\al X=0,
\end{equation}
where the covariant derivative acts with Christoffel's symbols on
Greek indices and the Lorentz connection on Latin ones.
\section{Canonical formulation                         \label{scanfo}}
The analysis of the model (\ref{efoact}) turns out to be much
simpler in the canonical formulation. Moreover the canonical
formulation allows one to write down a general solution to the
equations of motion in the matterless case without any gauge
fixing, the momenta being arbitrary functions parameterizing a
solution. The Hamiltonian structure of the quadratic model was
first analysed in the conformal gauge \cite{Katana89A}.

The canonical momenta for $\om_\al$, $e_\al{}^a$, and $X$ are given by
\begin{align}                                           \label{elcmze}
  \pi^0&=\frac{\pl(L_G+L_X)}{\pl(\pl_0\om_0)}=0,
\\                                                      \label{elcmfi}
  \pi^1&=\frac{\pl(L_G+L_X)}{\pl(\pl_0\om_1)}=\pi,
\\                                                      \label{ezwmze}
  p^0{}_a&=\frac{\pl(L_G+L_X)}{\pl(\pl_0e_0{}^a)}=0,
\\                                                      \label{ezwmfi}
  p^1{}_a&=\frac{\pl(L_G+L_X)}{\pl(\pl_0e_1{}^a)}=p_a,
\\                                                      \label{estcom}
  P&=\frac{\pl(L_G+L_X)}{\pl(\pl_0X)}
  =\rho\frac{g_{11}}e\pl_0X-\rho\frac{g_{01}}e\pl_1X.
\end{align}
The last momentum has dimension $[P]=1$. (For definition of dimensions
of the fields see Appendix.)
We see that the gravity Lagrangian (\ref{egrato}) is already written
in the canonical form. The  Hamiltonian for the whole system is given by
\begin{equation}                                        \label{ehamsd}
  H=\int\! d\s(\om_0 G+e_0{}^aG_a),
\end{equation}
where
\begin{align}                                           \label{escone}
  G&=-\pl_1\pi+p_a\ve^a{}_be_1{}^b,
\\                                                      \label{esctwo}
  G_a&=-\pl_1p_a-\om_1 p_b\ve^b{}_a+e_1{}^b\ve_{ab}
  \left(\frac12p^cp_cU+V-\frac\rho2m^2X^2\right)
\\                                                         \nonumber
  &+\frac{e_1{}^b\ve_{ab}}{g_{11}}
  \left(\frac1{2\rho}P^2+\frac\rho2\pl_1X^2\right)
  +\frac{e_{1a}}{g_{11}}P\pl_1X.
\end{align}
The functions $G$ and $G_a$ are a Lorentz scalar and vector,
respectively. Note that Hamiltonian (\ref{ehamsd}) for polynomial
$U(\pi)$ and $V(\pi)$ is polynomial in the fields in the absence of scalars.
Addition of a scalar field makes these functions nonpolynomial because of
the denominator $g_{11}$ and possible nonminimal coupling $\rho(\pi)$.

The equal time Poisson brackets are defined as usual
\begin{equation}                                        \label{epobrd}
\begin{split}
  \left\{e_1{}^a,p^\prime_b\right\}&=\dl^a_b\dl(\s-\s'),
\\
  \left\{\om_1,\pi^\prime\right\}&=\dl(\s-\s'),
\\
  \left\{X,P^\prime\right\}&=\dl(\s-\s'),
\end{split}
\end{equation}
where a prime over a function means that it is taken at a point $\s'$.
Computing the evolution of primary constraints (\ref{elcmze}) and
(\ref{ezwmze}) one gets the secondary constraints
\begin{align}                                        \label{esecon}
  \pl_0\pi^0&=\{p^0,H\}=-G=0,
\\                                                      \label{esectw}
  \pl_0p^0{}_a&=\{p^0{}_a,H\}=-G_a=0.
\end{align}
Thus the Hamiltonian (\ref{ehamsd}) is given by a linear combination of
secondary constraints. The secondary constraints form a closed algebra
\begin{align}                                        \label{efipbr}
  \{G_a,G'_b\}&=\ve_{ab}\left[Up^cG_c+
  \left(\frac12p^cp_cU'+V'-\frac{\rho'}{e\rho}L_X\right)G\right]\dl,
\\                                                      \label{esepbr}
  \{G_a,G'\}&=\ve_a{}^bG_b\dl,
\\                                                      \label{ethpbr}
  \{G,G'\}&=0,
\end{align}
where $U'$, $V'$, and $\rho'$ denote derivatives with respect to the
argument $\pi$, $\dl=\dl(\s-\s')$, and $L_X$ is the Lagrangian for
scalars (\ref{elagst}) expressed through canonical variables
$$
  L_X=-\frac12e\left(-\frac1{\rho g_{11}}P^2+\frac\rho{g_{11}}\pl_1X^2
  -m^2X^2\right).
$$
The Poisson brackets of the secondary constraints with the primary ones
vanish identically. Thus the model with the action (\ref{efoact})
possesses six first class constraints, and the Hamiltonian (\ref{ehamsd})
equals the linear combination of three secondary first class constraints.

At this point we are able to count
the number of physical continuous degrees of freedom. Suppose we have $D$
scalars. The gravity part in the second order form is described by the
zweibein (four components) and the Lorentz connection (two components).
Thus the total number of physical degrees of freedom equals $D+4+2-6$$=D$.
If we did not add the gravity part to scalars then the number of physical
degrees of freedom would be $D-2$ due to the symmetry of the Lagrangian
(\ref{elagst}) under general coordinate transformations.

An interesting point is that the Poisson bracket algebra
(\ref{efipbr})--(\ref{ethpbr}) closes with $\dl$-functions but not
with its derivatives. The "structure functions" depend not only on the
fields but on the functions $U(\pi)$, $V(\pi)$, and $\rho(\pi)$
entering the Lagrangian. In this sense the Poisson bracket algebra
depends on the dynamics of the fields. Note that in the usual canonical
formulation of a gravity model possessing invariance under general
coordinate transformation the constraints algebra closes with the first
derivatives of $\dl$-functions.

The Poisson bracket algebra (\ref{efipbr})--(\ref{ethpbr}) has a
(not invariant) subalgebra generated by the
constraint with a vector index
\begin{equation}                                        \label{evecoa}
  \widetilde G_a=G_a+k_a\om_1 G.
\end{equation}
Here for brevity we introduced a light-like vector
\begin{equation}                                        \label{elivec}
  k_a=\frac{e_{1a}+e_1{}^b\ve_{ba}}{g_{11}},~~~~~~k^ak_a=0,
\end{equation}
with the components
$$
  k_0=k_1=\frac{e_{10}+e_{11}}{g_{11}}.
$$
Straightforward calculations yield
\begin{align}                                           \label{einsal}
  \{\widetilde G_a,\widetilde G'_b\}&=
  \ve_{ab}(Up^c+\om_1 k^c)\widetilde G_c\dl,
\\                                                      \label{einsua}
  \{\widetilde G_a,G'\}&=\ve_a{}^b\widetilde G_b\dl-k_aG\dl'.
\end{align}
The algebra of $\widetilde G_a$ is related to the conformal algebra
generated by two scalar (with respect to Lorentz rotations) constraints
\begin{align}                                           \label{ehconf}
  H_0&=-e_1{}^a\ve_a{}^b\widetilde G_b=-e_1{}^a\ve_a{}^bG_b+\om_1 G,
\\                                                      \label{ehcont}
  H_1&=e_1{}^a\widetilde G_a=e_1{}^aG_a+\om_1 G.
\end{align}
The constraints $H_1$ and $H_0$ are Lorentz invariant projections of
the vector constraint $\widetilde G_a$ on the directions parallel and
perpendicular to the vector $e_1{}^a$, respectively. Straightforward
calculations show that the new set of constraints satisfy the following
algebra \cite{Katana94}
\begin{align}                                           \label{efirpb}
  \{H_0,H'_0\}&=-(H_1+H'_1)\dl',
\\                                                      \label{esecpb}
  \{H_0,H'_1\}&=-(H_0+H'_0)\dl',
\\                                                      \label{ethipb}
  \{H_1,H'_1\}&=-(H_1+H'_1)\dl',
\\                                                      \label{eforpb}
  \{H_0,G'\}&=\{H_1,G'\}=-G\dl',
\\                                                      \label{efifpb}
  \{G,G'\}&=0,
\end{align}
where
$$
  \dl'=\frac\pl{\pl\s'}\dl(\s'-\s).
$$
The constraints $H_0$ and $H_1$ form the well known conformal
algebra. The total algebra is the "semidirect product" of the conformal
algebra with an invariant abelian subalgebra generated by $G$ and
corresponding to local Lorentz rotations.

The inverse transformations to (\ref{ehconf}), (\ref{ehcont}) appear
as
\begin{align}                                           \label{eintro}
  G_0&=\frac1{g_{11}}\left[-e_1{}^1(H_0-\om_1 G)+e_1{}^0(H_1-\om_1 G)\right],
\\                                                      \label{eintgs}
  G_1&=\frac1{g_{11}}\left[e_1{}^0(H_0-\om_1 G)-e_1{}^1(H_1-\om_1 G)\right].
\end{align}

For later use we write down explicitly expressions for the constraints
$H_0$ and $H_1$ in terms of the canonical variables
\begin{align}                                          \nonumber
  H_0&=-\pl_1p_a\ve^a{}_be_1{}^b-\om_1 p_ae_1{}^a
  +\om_1(-\pl_1\pi+p_a\ve^a{}_be_1{}^b)
\\
  &-g_{11}\left(\frac12 p^ap_aU+V-\frac\rho2 m^2X^2\right)
  -\frac1{2\rho}P^2-\frac\rho2\pl_1X^2,
\\
  H_1&=-e_1{}^a\pl_1p_a-\om_1\pl_1\pi+P\pl_1X.
\end{align}
Note that the third term in $H_0$ is proportional to the constraint
$G$ (\ref{escone}).
\section{The canonical transformation                  \label{scantr}}
At the classical level two models related by a canonical transformation
are equivalent. At the quantum level this property is not valid
in general: There are canonical transformations resulting in different
quantum models. Special care must be given to nonlinear canonical
transformations. This means that in the canonical quantization the
correct choice of canonical variables is of primary importance.
Nobody knows the correct choice of variables because gravity is not
yet quantized. Therefore one is free to choose any set of
canonical variables if it leads to a simpler quantum model. Although
we do not consider quantization of the model in the present paper the
canonical variables introduced in this section are essential for the
solution of the constraints in Section \ref{scosol} already at the
classical level.

In this section we make the canonical transformation $e_1{}^a,p_a$
$\rightarrow$ $q,q_\perp,p,p_\perp$ which explicitly separates the
Lorentz angle and simplifies many formulas \cite{Katana94}. Consider
a generating functional depending on old coordinates and new momenta
\begin{equation}                                        \label{egecfu}
  F=\frac12\int\! d\s\left(p\,\ln|g_{11}|
  +p_\perp\ln\left|
  \frac{e_1{}^0+e_1{}^1}{e_1{}^0-e_1{}^1}\right|\right).
\end{equation}
Varying it with respect to the old coordinates one obtains the
relation between old and new momenta
\begin{equation}                                        \label{eolnep}
  p_a=p\frac{e_{1a}}{g_{11}}+p_\perp\frac{e_1{}^b\ve_{ba}}{g_{11}}.
\end{equation}
It shows that
$$
  p=e_1{}^ap_a,~~~~p_\perp=p_a\ve^a{}_be_1{}^b,
$$
that is, $p$ and $p_\perp$ are projections of the momentum $p_a$ on
the vector $e_1{}^a$ and the perpendicular direction.
Variation of the generating functional (\ref{egecfu}) with respect to
the momenta yields the relation between the coordinates
\begin{equation}                                        \label{ecotra}
\begin{split}
  e_1{}^0&=e^q\sh q_\perp,
\\
  e_1{}^1&=e^q\ch q_\perp.
\end{split}
\end{equation}
To drop the modulus signs in Eq.~(\ref{egecfu}) we assume for
definiteness that
\begin{equation}                                        \label{ecozwe}
  e_1{}^1>0~~~~\text{and}~~~~e_1{}^1>e_1{}^0.
\end{equation}
The space component of the metric equals to
$$
  g_{11}=-e^{2q}
$$
and is always negative.

We see that the coordinate $q_\perp$ coincides with the Lorentz
angle while $q$ parameterizes the length of the vector $e_1{}^a$.
The square of the momentum is
\begin{equation}                                           \label{emomsq}
  p^ap_a=(p_\perp^2-p^2)e^{-2q}.
\end{equation}

The constraints in new variables have the form
\begin{align}                                          \nonumber
  H_0&=-\pl_1p_\perp+p\pl_1q_\perp+p_\perp\pl_1q-\om_1 p
  +\om_1(-\pl_1\pi+p_\perp)
\\                                                      \label{efivco}
  &+\frac12(p_\perp^2-p^2)U+e^{2q}\left(V-\frac\rho2m^2X^2\right)
  -\frac1{2\rho}P^2-\frac\rho2\pl_1X^2,
\\                                                      \label{esevco}
  H_1&=-\pl_1p-\om_1\pl_1\pi+p\pl_1q+p_\perp\pl_1q_\perp+P\pl_1X,
\\                                                      \label{ethlco}
  G&=-\pl_1\pi+p_\perp.
\end{align}
For $U=\const$ the quadratic part in the momenta
$\frac12(p_\perp^2-p^2)$ corresponding to the gravitational degrees
of freedom has the same form as for two free particles with indefinite
signature. It means that the canonical transformation pushes the
nonpolynomiality from the kinetic term to the potential. Quantizing
the variables $q,p$ and $q_\perp,p_\perp$ one may construct the
Fock space representation without referring to a Lorentzian background
\cite{Katana94}. We postpone solution of the constraints
(\ref{efivco})--(\ref{ethlco}) to Section \ref{scosol}.
\section{Dilatonization                                \label{sdilto}}
If the Lorentz connection does not enter the matter Lagrangian then
it can be excluded from the model using its equations of motion.
As a consequence one obtains the generalized dilaton model, the
dilaton field coinciding with the momentum $\pi(x)$ conjugate to the
space component of the Lorentz connection $\om_1$. The equivalence
between gravity with torsion and dilaton gravity was first proved in
\cite{KaKuLi96}. There the transformation of variables included the
Weyl transformation of the metric and therefore did not preserve global
structure of solutions to the equations of motion \cite{KaKuLi96,KaKuLi97}.
Here we follow the improved procedure which does not change global
properties of space-time \cite{KuLiVa97,KumTie98}.

First, let us note the identity valid in two dimensions
\begin{equation}                                        \label{eidtwd}
  eR=e\widetilde R+\pl_\al(e\ve^\al{}_\bt T^{*\bt}),
\end{equation}
where $\widetilde R$ is the scalar curvature constructed from the
Christoffel symbols corresponding to zero torsion. After integration
by parts the gravity Lagrangian (\ref{egrato}) takes the form
\begin{equation}                                        \label{egrlas}
  L_G=-\frac12e\pi\widetilde R+\frac12\pl_\al\pi\ve^\al{}_a\widehat T^{*a}
      -\frac12p_a\widehat T^{*a}-e\left(\frac12p_ap^aU+V\right).
\end{equation}
We see that the Lorentz connection enters the Lagrangian only through
the torsion components $\widehat T^{*a}$, and they are related by the
invertible algebraic equation
\begin{equation}                                        \label{elocto}
  \om_\al=e_{\al a}\left(\frac12 T^{*a}-\ve^{\bt\g}\pl_\bt e_\g{}^a\right).
\end{equation}
Therefore instead of the Lorentz connection one can choose the torsion
components $T^{*a}$ as independent variables. Solving their algebraic
equations of motion one arrives at the generalized dilaton model
\begin{equation}                                        \label{edilmo}
  L_D=-\frac12e\pi\widetilde R+\frac12e\pl\pi^2 U-eV,
\end{equation}
depending on two arbitrary functions $U(\pi)$ and $V(\pi)$. Here we used
the obvious abbreviation $\pl\pi^2$$=g^{\al\bt}\pl_\al\pi\pl_\bt\pi$.
Thus, if the matter Lagrangian does not contain the Lorentz connection
then the dilaton model is equivalent to the two-dimensional gravity with
torsion. In fact, we have proved that if the original variables $e_\al{}^a$,
$\om_\al$, $p_a$, $\pi$, and $X$ satisfy equations of motion
(\ref{eqmopi})--(\ref{eqmozx}), then the variables $g_{\al\bt}$, $\pi$, and
$X$ satisfy equations of motion following from the Lagrangian $L_D+L_X$.
The inverse statement is as follows. Let $g_{\al\bt}$, $\pi$, and $X$
satisfy equations of motion for $L_D+L_X$. Solve the algebraic equation
$e_\al{}^a e_\bt{}^b\eta_{ab}$$=g_{\al\bt}$ for zweibein (a solution is
unique up to a local Lorentz rotations), construct $p_a$ using Equation
(\ref{eqmolc}), and solve equations (\ref{eqmotp}) and (\ref{elocto})
for $T^{*a}$ and $\om_\al$, respectively. Then the original equations
of motion (\ref{eqmopi})--(\ref{eqmozx}) will be satisfied.

The equivalence is a global one because the transformation of
variables is nondegenerate (\ref{elocto}) as far as nondegenerate is
the zweibein. This equivalence yields the geometric meaning for the
dilaton field: It is the momentum conjugate to the space component
of the Lorentz connection $\om_1$.

Variation of the action for the Lagrangian (\ref{edilmo}) with respect
to the metric and the dilaton field yields the equations of motion
\begin{align}                                           \label{emdilf}
  \frac1e\frac{\dl L_D}{\dl g_{\al\bt}}:&&
  -\frac12(g^{\al\bt}\widetilde\square\pi
  -\widetilde\nb^\al\widetilde\nb^\bt\pi)
  -\frac12\widetilde\nb^\al\pi\widetilde\nb^\bt\pi U+\frac12g^{\al\bt}
  \left(\frac12\pl\pi^2U-V\right)&=0,
\\                                                      \label{emdils}
  \frac1e\frac{\dl L_D}{\dl\pi}:&&
  -\frac12R-\widetilde\square\pi U-\frac12\pl\pi^2U'-V'&=0,
\end{align}
where $\widetilde\square=g^{\al\bt}\widetilde\nb_\al\widetilde\nb_\bt$
is the Laplace--Beltrami operator. This system of equations was solved in
the conformal gauge in \cite{Filipp96}, but we are not aware how to solve
directly this system of nonlinear equations of motion for arbitrary
functions $U$ and $V$ without gauge fixing. In the next section we
write down a general solution to the equivalent two-dimensional gravity
with torsion we started with in an arbitrary coordinate system. The
latter model turns out to be simpler,
and a general solution will be written even without gauge fixing.
\section{A general solution without matter             \label{sintmo}}
The important feature of two-dimensional gravity with torsion described
by the Lagrangian (\ref{egrato}) alone is its integrability. It has a
long history. First the quadratic model was solved in the conformal
\cite{Katana89B,Katana90} and light-cone \cite{KumSch92A} gauge. In
\cite{Solodu93A,MiGrObTrHe93,SchStr94A} a solution for the quadratic model
was in fact obtained without gauge fixing. Afterwards this solution was
clarified and generalized in the papers \cite{KumWid95,KumTie98}.
In this section we summarize all approaches and write a general local
solution of the equations of
motion for arbitrary functions $U$ and $V$ in the absence of matter
fields. It is naturally written in the canonical formulation.
This solution has one Killing vector field, and using the conformal
blocks technique one is able to construct easily all global (maximally
extended along extremals) solutions of two-dimensional gravity with
torsion or, equivalently, dilaton gravity for arbitrary given functions
$U$ and $V$. In this section we set $X=0$.
\subsection{Local solution                             \label{slosot}}
The integration of the equations of motion is most easily performed
for the light cone components of the vectors in the tangent space
\begin{equation}                                        \label{elicoc}
  p_\pm=\frac1{\sqrt 2}(p_0\pm p_1),~~~~~~
  e_1{}^\pm=\frac1{\sqrt 2}(e_1{}^0\pm e_1{}^1).
\end{equation}
The Lorentz metric and antisymmetric tensor for the tangent space
indices $a=\{+,-\}$ become
\begin{equation}                                        \label{elcmea}
  \et_{\pm\pm}=\left(\begin{array}{cc}0&1\\1&0\end{array}\right),~~~~~~
  \ve_{\pm\pm}=\left(\begin{array}{cc}0&-1\\1&0\end{array}\right).
\end{equation}
The raising and lowering of the light cone indices is performed according
to the rules $p^+=p_-$ and $p^-=p_+$. Then equations of motion
(\ref{eqmopi})--(\ref{eqmozw}) take the form
\begin{align}                                           \label{elcepi}
  -\frac12R-p_+p_-U'-V'&=0,
\\                                                      \label{elcpmi}
  -\frac12T^{*+}-p^+U&=0,
\\                                                      \label{elcppl}
  -\frac12T^{*-}-p^-U&=0,
\\                                                      \label{elcloc}
  \pl_\al\pi-e_\al{}^+p_++e_\al{}^-p_-&=0,
\\                                                      \label{elczwm}
  \pl_\al p_++\om_\al p_++e_\al{}^-(p_+p_-U+V)&=0,
\\                                                      \label{elczwp}
  \pl_\al p_--\om_\al p_--e_\al{}^+(p_+p_-U+V)&=0,
\end{align}
where
\begin{equation}                                        \label{etotrl}
  T^{*\pm}=2\ve^{\al\bt}(\pl_\al\mp\om_\al)e_\bt{}^\pm.
\end{equation}

Any local solution to these equations for $U\ne0$ and $V\ne0$ belongs to
one of the two classes. The first class corresponds to $p_-=0$ or,
equivalently, $p_+=0$. For $p_-=0$ the dilaton field is equal to the
constant, $\pi=\const$, defined by the algebraic equation
\begin{equation}                                        \label{econpi}
  V(\pi)=0
\end{equation}
as the consequence of Equation (\ref{elczwp}). Then $p_+=0$ due to
Equation (\ref{elcloc}), and Equation (\ref{elczwm}) is trivially
satisfied. The remaining Equations (\ref{elcepi})--(\ref{elcppl}) show
that the corresponding space-time is of zero torsion and constant
curvature
\begin{equation}                                        \label{eztcoc}
  T^{*\pm}=0,~~~~~~R=-2V'(\pi)=\const.
\end{equation}
In this way the constant curvature gravity model [1-3]
forms one class of solutions of two-dimensional gravity with torsion.

The second class of solutions, $p_-\ne0$ or $p_+\ne0$, corresponds to
nonzero torsion. For definiteness we assume that $p_-\ne0$ in some
domain. The case $p_+\ne0$ is treated in a similar way.
Integrability of the equations of motion relies on two nontrivial
observations. The first one is the existence of an integral of motion
\begin{equation}                                        \label{eintmo}
  A=p_+p_-e^{-Q}-W=\frac12(p_\perp^2-p^2)e^{-2q-Q}-W=\const,
\end{equation}
where $Q$ and $W$ are primitives
\begin{equation}                                        \label{eqwfun}
  Q(\pi)=\int^\pi\!dsU(s),~~~~~~W(\pi)=\int^\pi\!dsV(s)e^{-Q(s)}.
\end{equation}
They have dimensions
$$
  [Q]=1,~~~~[W]=[A]=l^{-2},
$$
as the consequence of (\ref{edimuv}) and (\ref{edippi}). The constants
of integration in Equations (\ref{eqwfun}) do not matter because $A$
by itself is a constant on the equations of motion
$$
  \pl_\al A=0.
$$
This equality is proved by direct calculations using equations of
motion (\ref{elcloc})--(\ref{elczwp}).
Note that this integral of motion depends only on the momenta and its
conservation is the consequence of eqs.\ (\ref{elcloc})--(\ref{elczwp})
only. Its existence will be clarified in the next section.
This integral of motion was independently found in dilaton gravity
\cite{LoGeKu94,GeKuLo95}.

The second important observation is that the form
\begin{equation}                                        \label{exacfo}
  dx^\al f_\al=\sqrt\kappa\frac{dx^\al e_\al{}^+}{p_-}e^Q
\end{equation}
is closed on the equations of motion. Here we introduced a dimensionfull
gravitational constant $[\kappa]=l^{-2}$ in order for $f_\al$ to be
dimensionless $[f_\al]=1$. To prove the closedness of the form
(\ref{exacfo}) one has to verify that the expression
$$
  \hat\ve^{\al\bt}\pl_\al f_\bt=0
$$
vanishes identically when equations of motion (\ref{elcpmi}),
(\ref{elcloc}), and (\ref{elczwp}) are satisfied. It means that at least
locally the one form (\ref{exacfo}) may be written as the gradient
of some scalar function
\begin{equation}                                        \label{exonfo}
  f_\al=\pl_\al f.
\end{equation}
Afterwards a general solution to the equations of motion
(\ref{elcepi})--(\ref{elczwp}) is written immediately
\begin{align}                                           \label{egensf}
  e_\al{}^+ &=\frac1{\sqrt\kappa}p_-e^{-Q}\pl_\al f,
\\                                                      \label{egenss}
  e_\al{}^- &=\frac1{p_-}\left[\frac1{\sqrt\kappa}(A+W)\pl_\al f
  -\pl_\al\pi\right],
\\                                                      \label{egenst}
  \om_\al &=\frac{\pl_\al p_-}{p_-}
           -\frac1{\sqrt\kappa}\left[(A+W)U+e^{-Q}V\right]\pl_\al f,
\\                                                      \label{egenfo}
  p_+ &=\frac1{p_-}(A+W)e^Q.
\end{align}
Here Equation (\ref{egensf}) is the consequence of (\ref{exacfo}) and
(\ref{exonfo}).
Equation (\ref{egenss}) follows from (\ref{elcloc}) and (\ref{eintmo}).
Equation (\ref{egenst}) is the consequence of (\ref{elczwp}) and
(\ref{eintmo}). Equation (\ref{egenfo}) is a simple rewriting of
(\ref{eintmo}). The next step of the proof is to verify that the
remaining equations of motion (\ref{elcepi})--(\ref{elcppl}) and
(\ref{elczwm}) are satisfied identically for arbitrary functions
$f,\pi$, and $p_-$. This can be checked by direct calculations,
and is the consequence of the linear dependence of the equations of
motion given by (\ref{edeplo}) and (\ref{edegco}). Note that the
solution (\ref{egensf})--(\ref{egenfo}) was obtained without any
gauge fixing and contains three arbitrary functions $f,\pi$ and $p_-$.
The first two functions correspond to the invariance of the model
under general coordinate transformations. They must satisfy the
restriction
\begin{equation}                                        \label{eresfp}
  d\pi\wedge df\ne0~~~~\text{or}~~~~\ve^{\al\bt}\pl_\al\pi\pl_\bt f\ne0.
\end{equation}
The third function $p_-$ corresponds to a Lorentz rotation and must be
strictly positive $p_->0$ or negative $p_-<0$.

The second class of solutions describes surfaces with nonzero torsion
and nonconstant scalar curvature which are most easily calculated using
Equations (\ref{elcepi})--(\ref{elcppl}). The corresponding metric is
\begin{equation}                                        \label{emetsc}
  ds^2=2e_\al{}^+e_\bt{}^-dx^\al dx^\bt
  =2e^{-Q}\left[\frac1\kappa(A+W)df^2-\frac1{\sqrt\kappa}dfd\pi\right].
\end{equation}
We see that the scalar functions $f$ and $\pi$ are the
Eddington--Finkelstein coordinates \cite{Edding24,Finkel58}, the
dilaton field being the lightlike one.

The components of the metric (\ref{emetsc}) depend only on the dilaton
field $\pi$, and hence the solution has one Killing vector field
\begin{equation}                                        \label{ekilse}
  K=\frac1{\sqrt\kappa}\frac{\pl}{\pl f}.
\end{equation}
Its square is
\begin{equation}                                        \label{elenkv}
  K^2=\frac2{\kappa^2}e^{-Q}(A+W),
\end{equation}
and may be timelike or spacelike depending on the sign of $A+W$. Along
horizons defined by the equation $A+W=0$ the Killing vector is lightlike.

A general solution (\ref{egensf})--(\ref{egenfo}) in new canonical
variables introduced in Section~\ref{scantr} takes the form
\begin{align}                                           \label{egespe}
  p&=\frac2{\sqrt\kappa}(A+W)\pl_1f-\pl_1\pi,
\\                                                      \label{egespp}
  p_\perp&=\pl_1\pi,
\\                                                      \label{egesqu}
  e^{2q}&=\frac2\kappa\pl_1f[\sqrt\kappa\pl_1\pi-(A+W)\pl_1f]e^{-Q},
\\                                                      \label{egesqp}
  e^{2q_\perp}&=\frac{p_-^2\pl_1f}
  {\sqrt\kappa\pl_1\pi-(A+W)\pl_1f}e^{-Q}.
\end{align}
It is valid only for these functions $\pi$ and $f$ which provide
positive definiteness of the right hand sides of (\ref{egesqu}) and
(\ref{egesqp}) according to our assumption (\ref{ecozwe}).

The above solution is obtained under the restriction $p_-\ne0$.
In a domain with $p_+\ne0$ we still have the integral of motion
(\ref{eqforb}) but the form (\ref{exacfo}) must be replaced because we
admit zero of $p_-$. One may check that the form
$$
  df=\sqrt\kappa\frac{dx^\al e_\al{}^-}{p_+}e^Q
$$
is closed on the equations of motion (\ref{elcppl}), (\ref{elcloc}), and
(\ref{elczwm}). A similar procedure results in a general
solution to the equations of motion
\begin{align}                                           \label{egessf}
  e_\al{}^- &=\frac1{\sqrt\kappa}p_+e^{-Q}\pl_\al f,
\\                                                      \label{egesss}
  e_\al{}^+ &=\frac1{p_+}\left[\frac1{\sqrt\kappa}(A+W)\pl_\al f
  +\pl_\al\pi\right],
\\                                                      \label{egesst}
  \om_\al &=-\frac{\pl_\al p_+}{p_+}
  -\frac1{\sqrt\kappa}\left[(A+W)U+e^{-Q}V\right]\pl_\al f,
\\                                                      \label{egesfo}
  p_- &=\frac1{p_+}(A+W)e^Q,
\end{align}
where $p_+$ is considered as an arbitrary function parameterizing
the solution. The corresponding metric differs from (\ref{emetsc})
by the sign before the second term
\begin{equation}                                        \label{emessc}
  ds^2=2e^{-Q}
  \left[\frac1\kappa(A+W)df^2+\frac1{\sqrt\kappa}dfd\pi\right].
\end{equation}
This difference may be eliminated by the redefinition $f\rightarrow-f$.

At the end of this section we make a short comment on the conserved
quantity A (\ref{eintmo}). For the Schwarzschild solution
it equals a mass of the black hole up
to a constant factor (see Section \ref{sredgr}). The equation
\begin{equation}                                        \label{eacons}
  A=\const
\end{equation}
is a first class constraint of the model. One may check that its
space derivative is expressed in terms of the secondary constraints
\begin{equation}                                        \label{eatesc}
  \pl_1 A=e^{-Q}\left[-p^aG_a+\left(\frac12p^ap_aU+V\right)G\right].
\end{equation}
The Poisson brackets of the constraint (\ref{eacons}) with the secondary
constraints are zero,
$$
  \{A,G'_a\}=\{A,G'\}=0,
$$
and thus $A$ belongs to the center of the algebra
(\ref{efipbr})--(\ref{ethpbr}).
Straightforward calculations yield the following Poisson brackets of
this constraint with the other constraints:
{\allowdisplaybreaks
\begin{align*}                                  
  &\{A,\widetilde G'_a\}=
  \left[k^cp_c\om_1+\frac12p^cp_cU+V\right]k_ae^{-Q}G\dl,
\\                                               
  &\{A,\widetilde G'_0\}=\{A,\widetilde G'_1\}
  =\left[(p_\perp-p)e^{-2q}\om_1
  +\frac12(p_\perp^2-p^2)e^{-2q}U+V\right]
  e^{-(q+q_\perp+Q)}G\dl,
\\                                                
  &\{A,H'_0\}=\left[-\frac{e_1{}^ap_a}{g_{11}}H_0
  -\frac{e_1{}^a\ve_a{}^bp_b}{g_{11}}H_1
  +\left(k^ap_a\om_1+\frac12p^ap_aU+V\right)G\right]e^{-Q}\dl,
\\                                                 
  &=\left[pe^{-2q}H_0-p_\perp e^{-2q}H_1
  +\left((p_\perp-p)e^{-2q}\om_1
  +\frac12(p_\perp^2-p^2)e^{-2q}U+V\right)G\right]e^{-Q}\dl,
\\                                                  
  &\{A,H'_1\}=\left[\frac{e_1{}^a\ve_a{}^bp_b}{g_{11}}H_0
  +\frac{e_1{}^ap_a}{g_{11}}H_1
  +\left(-k^ap_a\om_1+\frac12p^ap_aU+V\right)G\right]e^{-Q}\dl
\\                                                   
  &=\left[p_\perp e^{-2q}H_0-pe^{-2q}H_1
  +\left(-(p_\perp-p)e^{-2q}\om_1
  +\frac12(p_\perp^2-p^2)e^{-2q}U+V\right)G\right]e^{-Q}\dl.
\end{align*}}
Here we write the Poisson brackets in the coordinates both before and
after the canonical transformation for comparison. We see that the
constraint (\ref{eacons}) does not form a closed algebra with the
conformal constraints $H_0$ and $H_1$ alone.
\subsection{Local solution for a general Lagrangian    \label{sgenla}}
Local solution to the equations of motion for the gravity Lagrangian
found in the previous section relies on the existence of the
conserved quantity (\ref{eintmo}) and the closed form (\ref{exacfo}).
To clarify the appearance of the conservation law (\ref{eintmo}) we
consider a more general gravity Lagrangian
\begin{equation}                                        \label{egragl}
  L_G=-\frac12(\pi\widehat R+p_a\widehat T^{*a})-e\CU,
\end{equation}
where $\CU(\wp,\pi)$ is an arbitrary function of two scalar functions
$\wp=p^ap_a$ and $\pi$. For the linear function $\CU=\wp U/2+V$ in $\wp$
we recover the original gravity model (\ref{egrato}). In addition we clarify
the statement that equations of motion are integrable in this more general
case \cite{MiGrObTrHe93,Solodu94A,KloStr96A}.

Equations of motion for the Lagrangian (\ref{egragl}) in light-cone
coordinates take the form
\begin{align}                                           \label{eqmlpi}
  \frac{\dl S}{\dl\pi}&:&-&
  \hat\ve^{\al\bt}(\pl_\al\om_\bt+\CU,{}_\pi e_\al{}^+e_\bt{}^-)=0,
\\                                                      \label{eqmltp}
  \frac{\dl S}{\dl p_+}&:&-&\hat\ve^{\al\bt}
  (\pl_\al e_\bt{}^+-\om_\al e_\bt{}^++2p_-\CU,{}_\wp e_\al{}^+e_\bt{}^-)=0,
\\                                                      \label{eqmltm}
  \frac{\dl S}{\dl p_-}&:&-&\hat\ve^{\al\bt}
  (\pl_\al e_\bt{}^-+\om_\al e_\bt{}^-+2p_+\CU,{}_\wp e_\al{}^+e_\bt{}^-)=0,
\\                                                      \label{eqmllc}
  \hat\ve_{\bt\al}\frac{\dl S}{\dl\om_\bt}&:&&
  \pl_\al\pi-p_+e_\al{}^++p_-e_\al{}^-=0,
\\                                                      \label{eqmlzw}
  \hat\ve_{\bt\al}\frac{\dl S}{\dl e_\bt{}^+}&:&&\pl_\al p_++\om_\al p_+
  +e_\al{}^-\CU=0,
\\                                                      \label{eqmlzm}
  \hat\ve_{\bt\al}\frac{\dl S}{\dl e_\bt{}^-}&:&&\pl_\al p_--\om_\al p_-
  -e_\al{}^+\CU=0,
\end{align}
where
$$
  \CU,{}_\wp=\frac{\pl\CU}{\pl\wp},~~~~~~\CU,{}_\pi=\frac{\pl\CU}{\pl\pi}.
$$
We have nine equations for nine independent variables
$\pi,p_\pm,\om_\al,e_\al{}^\pm$. There are three linear relations between
equations of motion (\ref{edeplo}), (\ref{edegco}) due to the symmetry
under general coordinate transformations and local Lorentz rotations.
It means that to find a general solution to the equations of motion one
has to solve only six independent equations.

Equations (\ref{eqmlpi})--(\ref{eqmltm}) may be written in the form
\begin{align}                                           \label{egescr}
  -\frac12R-\CU,{}_\pi&=0,
\\                                                      \label{egetoc}
  -\frac12T^{\star\pm}-2p^\pm\CU,{}_\wp&=0.
\end{align}
where the scalar curvature and torsion are defined by equations
(\ref{escden}), (\ref{epttod}). If these equations have a unique solution
with respect to $\pi$ and $p^\pm$ then the Lagrangian (\ref{egragl}) is
nothing else then the Legendre transform of some function
\begin{equation}                                        \label{eletrl}
  L_G^{(2)}=eF(R,T^2)
\end{equation}
depending on the scalar curvature $R$ and the torsion squared term
$T^2=T^*_aT^{*a}$ with respect to three variables $-R/2$ and $-T^{*a}/2$.
The Lagrangian written in the first order form (\ref{egragl}) is more
general because we do not assume that a function $\CU(\wp,\pi)$ admits
the Legendre transformation.

Let us try to find the six unknowns $\om_\al,e_\al{}^\pm$ in terms of
the conjugate momenta considered as arbitrary functions. For fixed
index $\al$ Equations (\ref{eqmllc})--(\ref{eqmlzm}) constitute a set
of three linear algebraic inhomogeneous equations for $\om_\al,e_\al{}^\pm$.
Its determinant vanishes identically, and hence for any nontrivial solution
there must exist a relation between the momenta. To find it one may take
the linear combination of the equations
(\ref{eqmlzw})$p_-+$(\ref{eqmlzm})$p_+$ and using equation (\ref{eqmllc})
find
$$
  \pl_\al\wp-2\pl_\al\pi\CU=0.
$$
It means that momenta must satisfy the following ordinary differential
equation
\begin{equation}                                        \label{eordmo}
  \frac{d\wp}{d\pi}=2\CU(\wp,\pi).
\end{equation}
It is an integrability condition for Equations (\ref{eqmllc})--(\ref{eqmlzm}).
For reasonable functions $\CU$ a general solution of this equations exists
and depends on one arbitrary constant which is the integral of motion.
For linear $\CU$ it is given by Equation (\ref{eintmo}). Though the
integral of motion exists for arbitrary functions $\CU$, it is not enough
to integrate explicitly equations of motion in a general case.

The existence of this integral of motion means that only two momenta of
three are independent. Let it be $\pi$ and $p_-$. Afterwards $p_+$ is
to be found from a solution of Equation (\ref{eordmo}). Now we use two
Equations (\ref{eqmllc}), (\ref{eqmlzm}) to find
\begin{align}                                           \label{ealmin}
  e_\al{}^-&=\frac1{p_-}(-\pl_\al\pi+p_+e_\al{}^+),
\\                                                      \label{eomals}
  \om_\al&=\frac1{p_-}(\pl_\al p_--e_\al{}^+\CU).
\end{align}
To complete integration of the equations of motion one has to find
$e_\al{}^+$. These two components are to be found as a solution of
one of the Equations (\ref{eqmlpi})--(\ref{eqmltm}) or their linear
combination because algebraic Equations (\ref{eqmllc})--(\ref{eqmlzm})
are dependent. Taking the linear combination
(\ref{eqmltp})$p_++$(\ref{eqmltm})$p_-$ and using Equations
(\ref{eqmlzw}), (\ref{eqmlzm})
one gets the linear differential equation for $p_+e_\al{}^+$
\begin{equation}                                        \label{eqppep}
  \hat\ve^{\al\bt}\left[\pl_\al(p_+e_\bt{}^+)-2(\CU-\wp\CU,{}_\wp)
  \frac{\pl_\al\pi}\wp(p_+e_\bt{}^+)\right]=0.
\end{equation}
Without loss of generality we set
$$
  p_+e_\al{}^+=\wp e^{-Q}f_\al,
$$
where $f_\al$ is a one form and $Q(\wp,\pi)$ is some unknown scalar
function of two variables to be specified later. Then Equation
(\ref{eqppep}) takes the form
\begin{equation}                                        \label{eqftoc}
  \wp e^{-Q}\hat\ve^{\al\bt}[\pl_\al f_\bt-\pl_\al\pi f_\bt
  (2\CU Q,{}_\wp+Q,{}_\pi-2\CU,{}_\wp)]=0.
\end{equation}
Solution of this equation always exists for reasonable functions
$\pi$, $\CU$, and $Q$, but a solution cannot be written explicitly
in a general case. It can be exactly integrated in a particular case.
If we choose the function $Q$ satisfying the partial differential equation
\begin{equation}                                        \label{eqforb}
  2\CU Q,{}_\wp+Q,{}_\pi-2\CU,{}_\wp=0,
\end{equation}
then a one form $f_\al$ must be closed. It means that locally this one
form is exact (\ref{exonfo}). Hence we get a general solution
\begin{equation}                                        \label{expepl}
  e_\al{}^+=2p_-e^{-Q}\pl_\al f,
\end{equation}
where $f$ is an arbitrary function, and $Q$ is a solution of the partial
differential Equation (\ref{eqforb}). The rest of the equations of motion
are satisfied as the consequence of their linear dependence.

Equations for characteristics $\wp(\pi),Q(\wp(\pi),\pi)$ for Equation
(\ref{eqforb}) are
\begin{equation}                                        \label{eficha}
\begin{split}
  \frac{d\wp}{d\pi}&=2\CU,
\\
  \frac{dQ}{d\pi}&=2\frac{\pl\CU}{\pl\wp}.
\end{split}
\end{equation}
Note that the first Equation (\ref{eficha}) coincides with the
integrability condition (\ref{eordmo}).

Thus we get a general solution to the equations of motion for arbitrary
$\CU(\wp,\pi)$ without any gauge fixing. It is given by (\ref{ealmin}),
(\ref{eomals}), and (\ref{expepl}) where $Q$ is a solution of Equation
(\ref{eqforb}), and $p_+$ is to be found from Equation (\ref{eordmo}).
This solution depends on three arbitrary functions $\pi,f$, and $p_-$,
corresponding to general coordinate transformations and local Lorentz
rotations with the only restriction (\ref{eresfp}).

In this way the integration of the equations of motion is reduced to one
ordinary differential Equation (\ref{eordmo}) and one equation with
partial derivatives (\ref{eqforb}). It is hardly justified to say that
equations of motion are integrable because solution of the last equation
is known to exist for sufficiently smooth $\CU$ but cannot be written
in quadratures in a general case. It is an
interesting question to find functions $\CU$ for which solution of
Equation (\ref{eqforb}) can be written explicitly thus providing
integrable two-dimensional gravity models. In this paper we consider
only linear functions $\CU=\wp U/2+V$ in $\wp$ corresponding to gravity
models quadratic in torsion. These models are really integrable with a
solution for Equation (\ref{eqforb}) given by (\ref{eqwfun}).
Besides, we have the usual kinetic term for the dilaton field after
dilatonization which should be modified for more general Lagrangians.
\subsection{Global solutions                           \label{sglsot}}
The solution found in the preceeding sections is a local one. To give
a physical interpretation of the solution and to understand global
structure of the space-time one has to extend the solution along
extremals (geodesics). This can be done using the constructive conformal
block method developed for the Lorentzian \cite{Walker70,Katana93A,Katana00A}
and Euclidean \cite{Katana97} signature metric. An equivalent set of rules
in the Eddington--Finkelstein coordinates may be found in
\cite{KloStr96C,KloStr97}. To apply the conformal blocks technique
one has to rewrite local solutions
obtained in the previous section in the conformal form. This should be
done in every domain where the solution is defined because a global
solution is obtained by gluing all patches together.

Before doing this we write the metric (\ref{emetsc}) in the diagonal
gauge for comparison with the metric in the matterfull case.
In a domain with $A+W>0$ we leave the coordinate $\pi$ as it is and
transform the coordinate $f$ only
\begin{equation}                                        \label{etrdio}
  f=\frac12\tau+g(\pi),
\end{equation}
where the function $g(\pi)$ is defined by the equation
\begin{equation}                                        \label{edefgp}
  g'=\frac{\sqrt\kappa}{2(A+W)}.
\end{equation}
This transformation is nondegenerate at least between horizons defined by
the equation $A+W=0$. Then the metric (\ref{emetsc}) takes a diagonal form
\begin{equation}                                        \label{emaled}
  ds^2=\frac12e^{-Q}\left[\frac1\kappa(A+W)d\tau^2-\frac{d\pi^2}{A+W}\right].
\end{equation}
For $A+W>0$ coordinates $\tau$ and $\pi$ are timelike and spacelike,
respectively.

This metric can be easily rewritten in the conformal gauge suitable for
the global analysis. Introducing the space coordinate $\s$ defined
by the equation
$$
  \frac{d\pi}{d\s}=\pm\frac{A+W}{\sqrt\kappa},
$$
and keeping $\tau$ unchanged, one gets a conformally flat metric
\begin{equation}                                        \label{ecofso}
  ds^2=\frac1{2\kappa}e^{-Q}(A+W)(d\tau^2-d\s^2),~~~~~~A+W>0.
\end{equation}
Introducing the invariant variable $\hat q$ related to $\pi$ by ordinary
differential equation
\begin{equation}                                        \label{einhaq}
  \frac{d\hat q}{d\pi}=\frac1{\sqrt\kappa}e^{-Q}
\end{equation}
the solution (\ref{ecofso}) may be rewritten in the form
\begin{align}                                           \label{esocfl}
  ds^2&=N(\hat q)(d\tau^2-d\s^2),~~~~~~N(\hat q)>0,
\\ \intertext{where}                                    \label{ecocfa}
  \pm\frac{d\hat q}{d\s}&=N(\hat q),
\\ \intertext{and the conformal factor is}              \label{ecofad}
  N(\hat q)&=\frac1{2\kappa}e^{-Q}(A+W).
\end{align}

In a domain $A+W<0$ we introduce a space coordinate
$$
  f=\frac12\s+g(\pi),
$$
where $g(\pi)$ is defined by the same Equation (\ref{edefgp}). Then
solution (\ref{emetsc}) takes a diagonal form
$$
  ds^2=-\frac12e^{-Q}\left[\frac{d\pi^2}{A+W}-\frac1\kappa(A+W)d\s^2\right].
$$
Introducing the time coordinate
$$
  \frac{d\pi}{d\tau}=\pm\frac{A+W}{\sqrt\kappa},
$$
the metric takes a conformally flat form
$$
  ds^2=-\frac1{2\kappa}e^{-Q}(A+W)(d\tau^2-d\s^2),~~~~~~A+W<0.
$$
In terms of the invariant variable $\hat q$ (\ref{einhaq}) this metric
takes the form
\begin{align}                                        \label{ecofmh}
  ds^2&=-N(\hat q)(d\tau^2-d\s^2),~~~~~~N(\hat q)<0,
\\ \intertext{where the conformal factor is given by Equation
(\ref{ecofad}) and}                                   \nonumber
  \pm\frac{d\hat q}{d\tau}&=-N(\hat q).
\end{align}

Solutions in domains $A+W>0$ and $A+W<0$ may be united in the way
\begin{align*}
  ds^2&=\left|N(\hat q)\right|(d\tau^2-d\s^2),
\\
  \left|\frac{d\hat q}{d\z}\right|&=\pm N(\hat q),
\end{align*}
where $\z=\s$ and $\z=\tau$ for positive and negative values of the
conformal factor $N(\hat q)$, respectively.
Afterwards one may apply the conformal block technique
and construct global (maximally extended) solutions for the matterless
models for any functions $U$ and $V$. The method is straightforward and
allows one to construct Carter--Penrose diagrams by the analysis of
zero and singularities of the conformal factor $N(\hat q)$.
The method is reviewed and several examples are given in \cite{Katana00A}.
\section{Solution of the constraints                   \label{scosol}}
From now on we return to the matterfull case $X\ne0$. To construct the
effective action for scalars one has to solve the geometric
part of the equations of motion (\ref{eqmopi})--(\ref{eqmozw}) assuming
that a scalar field $X$ and its conjugate momenta $P$ are arbitrary
functions. This will be done in two steps. First, we solve three
constraints (\ref{efivco})--(\ref{ethlco}), which are equivalent to the
variational derivatives of the action with respect to the time components
of the zweibein and Lorentz connection $\dl S/\dl e_0{}^a$ and
$\dl S/\dl\om_0$. Afterwards we solve the rest of the geometric part
of the equations of motion.

The constraints $G$ and $G_a$ in the light cone coordinates are
\begin{align}                                           \label{egclco}
  G~&=-\pl_1\pi+e_1{}^+p_+-e_1{}^-p_-,
\\                                                      \label{egplic}
  G_\pm&=-\pl_1p_\pm\mp\om_1 p_\pm\mp e_1{}^\mp
  \left(p_+p_-U+V-\frac\rho2m^2X^2\right)
  \mp\frac\rho{4e_1{}^\pm}P_\mp^2,
\end{align}
where we introduced a shorthand notation
\begin{equation}                                           \label{edepxs}
  P_\pm=\frac1\rho P\pm\pl_1X.
\end{equation}
These constraints were solved in the absence of matter in the previous
section. To find a solution to the equations $G=G_\pm=0$ for the arbitrary
scalar field we use the canonical transformation introduced in Section
\ref{scantr} and use the equivalent set of constraints $G=H_0=H_1=0$.
Here we shall see the advantage of the canonical transformation.

To simplify the procedure we fix the gauge; that is, we fix a coordinate
system and local Lorentz rotations. This is done in a slightly unusual way.
Namely, we consider a wide class of gauges of the form
\begin{equation}                                        \label{esegac}
\begin{split}
  F_1&=\pi-\pi(\tau,\s)=0,
\\
  F_2&=p-p(\tau,\s)=0,
\\
  F_3&=q_\perp-q_\perp(\tau,\s)=0.
\end{split}
\end{equation}
They state that the two momenta $\pi,p$ and the Lorentz angle $q_\perp$
are given functions of space-time coordinates $\pi(\tau,\s)$,
$p(\tau,\s)$, and $q_\perp(\tau,\s)$. We call them gauge fixing functions.
The first two functions specify the coordinate system while the third one
fixes a local Lorentz rotation. The form of these functions will be
specified later to simplify the effective action for scalars. Let us
note that we do not fix the fields by themselves but their conjugate momenta.
This shows the strength of the canonical formulation. It is not easy to
imagine how one can fix derivatives of the fields keeping the original
fields as independent variables within the Lagrangian second order
approach. The class of gauges (\ref{esegac}) is very
wide until we specify the gauge fixing functions. In fact, we only
choose the variables which have to be fixed. It is enough to write
down a solution for all other geometrical variables in terms of scalars.

The constraints (\ref{efivco})--(\ref{ethlco}) contain only canonically
transformed space components of the geometric variables. Three of them
are fixed by gauge conditions (\ref{esegac}), and hence
we have three equations for three unknown functions $\om_1,q,p_\perp$.
The last constraint is trivially solved
\begin{equation}                                        \label{emoper}
  p_\perp=\pl_1\pi.
\end{equation}
Here we see the importance of the canonical transformation which simplified
greatly the Lorentz constraint (\ref{escone}).

Afterwards the two remaining constraints become
\begin{align}                                           \nonumber
  H_0=&-\pl_1^2\pi+p(\pl_1q_\perp-\om_1)+\pl_1\pi\pl_1q+
  \frac12YU+e^{2q}(V-\frac\rho2m^2X^2)
\\                                                      \label{egcfix}
  &-\frac1{2\rho}P^2-\frac\rho2\pl_1X^2=0,
\\                                                      \label{egmgfi}
  H_1=&-\pl_1p+\pl_1\pi(\pl_1q_\perp-\om_1)+p\pl_1q+P\pl_1X=0,
\end{align}
where we introduced a shorthand notation for the quadratic combination of
the gauge fixing functions
\begin{equation}                                        \label{edefyp}
  Y=\pl_1\pi^2-p^2,~~~~~~[Y]=l^{-2}.
\end{equation}
There are four different cases depending on the gauge fixing functions
\begin{align}                                            \label{ecasea}
  \text{Case {\bf A}}:~~~~&Y\ne0,~~\pl_1\pi\ne0,
\\
  \text{Case {\bf B}}:~~~~&Y\ne0,~~\pl_1\pi=0,
\\
  \text{Case {\bf C}}:~~~~&Y=0,~~\pl_1\pi\ne0.
\\
  \text{Case {\bf D}}:~~~~&Y=0,~~\pl_1\pi=0.
\end{align}
The first case describes a general situation, while the last three are
degenerate ones. The fourth possibility, which may be written in an
equivalent form $p=0$ and $\pl_1\pi=0$, restricts the scalars as the
consequence of the constraint $H_1=0$, and does not allow us to find
$\om_1$ from the constraints. In the last two cases
at least one of two light-cone components of torsion is zero.
Indeed, the square of the momenta (\ref{emomsq}) together with the
constraint (\ref{ethlco}) takes the form
\begin{equation}                                       \label{emosgp}
  p^ap_a=2p_+p_-=Ye^{-2q}.
\end{equation}
It means that for $Y=0$ at least one of the light cone components of
torsion must be zero as the consequence of Equation (\ref{eqmotp}) or
(\ref{elcpmi}), (\ref{elcppl})
which are not affected by the presence of scalars.

{\em The case {\bf A}}. $Y\ne0$, $\pl_1\pi\ne0$. 
For $\pl_1\pi\ne0$ Equation (\ref{egmgfi}) yields
\begin{equation}                                        \label{eqpero}
  \pl_1q_\perp-\om_1=\frac1{\pl_1\pi}(\pl_1p-p\pl_1q-P\pl_1X).
\end{equation}
Plugging this solution for $\om_1$ into the constraint (\ref{egcfix})
one gets the equation for $q$
\begin{align}                                           \nonumber
  H_0\pl_1\pi&=\pl_1qY+e^{2q}\left(V-\frac\rho2m^2X^2\right)\pl_1\pi
  -\frac12\pl_1Y+\frac12YU\pl_1\pi-pP\pl_1X
\\                                                      \label{eqforq}
  &-\pl_1\pi\left(\frac1{2\rho}P^2+\frac\rho2\pl_1X^2\right)=0.
\end{align}
Provided $Y\ne0$ this first order nonlinear ordinary differential
equation for $q$ has a general solution
\begin{equation}                                        \label{esolfq}
  2q=\ln\frac Y{2(A_m+W_m)}-Q+2T,
\end{equation}
where $Q$ is defined by the integral (\ref{eqwfun}) and two
integrals are introduced,
\begin{align}                                        \label{edefze}
  T(\tau,\s)&=-\int_\s^\infty\frac{d\s'}Y\left[pP\pl_1X+\pl_1\pi
  \left(\frac1{2\rho}P^2+\frac\rho2\pl_1X^2\right)\right],
\\                                                   \label{edefdw}
  W_m(\tau,\s)&=\int_0^\s d\s'\pl_1\pi
  \left(V-\frac\rho2m^2X^2\right)e^{-Q+2T}.
\end{align}
They have dimensions
$$
  [T]=1,~~~~~~[W_m]=l^{-2}.
$$
This solution depends on one arbitrary function of time $A_m(\tau)$
entering (\ref{esolfq}). The integrals (\ref{edefze}), (\ref{edefdw})
are assumed to be convergent. If not, then the upper and lower limits
of integration in (\ref{edefze}) and (\ref{edefdw}), respectively, have
to be redefined. The limits of integration are adopted for spherically
reduced gravity considered in Section \ref{sredgr}. The solution
(\ref{esolfq}) yields the solution for the space component of the metric
\begin{equation}                                        \label{esoqex}
  -g_{11}=e^{2q}=\frac Y{2(A_m+W_m)}e^{-Q+2T},
\end{equation}
This solution is valid on those patches were expression (\ref{esoqex}) is
positive. A similar solution may be written for negative values of the
right hand side of (\ref{esoqex}). In the matterless
case $T=0$, $W_m$ coincides with $W$ given by (\ref{eqwfun}), and
$A_m=A=\const$ as the consequence of the remaining equations of motion
(\ref{eqmopi})--(\ref{eqmozw}) (this statement will be proved in Section
\ref{sreeqz}).

Substitution of the solution (\ref{esoqex}) into (\ref{eqpero}) yields
the explicit expression for the space component of the Lorentz
connection
\begin{align}
  \om_1&=\pl_1q_\perp                                   \nonumber
  -\frac12\pl_1\ln\left|\frac{\pl_1\pi+p}{\pl_1\pi-p}\right|
  -\frac12pU-\frac{p\left(V-\frac\rho2m^2X^2\right)}{2(A_m+W_m)}e^{-Q+2T}
\\                                                      \label{esolos}
  &+\frac{\pl_1\pi}YP\pl_1X
  +\frac pY\left(\frac1{2\rho}P^2+\frac\rho2\pl_1X^2\right).
\end{align}
Together with Equations (\ref{emoper}) and (\ref{esolfq}) it
provides a general solution to the constraints in the case {\bf A}
(\ref{ecasea}).

{\em The case {\bf B}}. $Y\ne0$, $\pl_1\pi=0$. 
Now the constraints are
\begin{align}                                           \label{ehzcoz}
  H_0&=p(\pl_1q_\perp-\om_1)-\frac12p^2U+e^{2q}\left(V-\frac\rho2m^2X^2\right)
  -\frac1{2\rho}P^2-\frac\rho2\pl_1X^2=0,
\\                                                      \label{ehocos}
  H_1&=-\pl_1p+p\pl_1q+P\pl_1X=0.
\end{align}
First we solve the last constraint with respect to $q$
\begin{equation}                                        \label{esohoz}
  q=B(\tau)+\ln p-\int_0^\s\frac{d\s'}p P\pl_1X,
\end{equation}
where $B=B(\tau)$ is an arbitrary function of time. Afterwards the
constraint $H_0$ is solved with respect to $\om_1$
\begin{equation}                                        \label{esohfz}
  \om_1=\pl_1q_\perp-\frac1p\left(\frac12p^2U-p^2
  \left(V-\frac\rho2m^2X^2\right)e^{2B-2\int\!\frac{d\s}pP\pl_1X}
  +\frac1{2\rho}P^2+\frac\rho2\pl_1X^2\right).
\end{equation}

Though the solution of the constraints in this case differ from Case
{\bf A}, Equations (\ref{esohoz}), (\ref{esohfz}) coincide formally with
Equations (\ref{esolfq}) and (\ref{esolos}) for $\pl_1\pi=0$ with a
suitable redefinition of $B(\tau)$. Therefore Case {\bf B} may be
considered as a subcase of {\bf A}.

{\em The case {\bf C}}. $Y=0$, $\pl_1\pi\ne0$. 
In this case equations for $q$ and $\om_1$ are purely algebraic. The
condition $Y=0$ yields
\begin{equation}                                        \label{epmrup}
  p=\pm\pl_1\pi.
\end{equation}
For two possible choices of signs the difference $H_0-H_1$ and the sum
$H_0+H_1$ of the constraints (\ref{egcfix}), (\ref{egmgfi}) immediately
yield
\begin{equation}                                        \label{espcmc}
  -g_{11}=e^{2q}=\frac{\rho P_\pm^2}{2\left(V-\frac\rho2 m^2X^2\right)},
\end{equation}
where $P_\pm$ is defined by (\ref{edepxs}).
Afterwards the constraint $H_1$ yields a solution for $\om_1$
\begin{equation}                                        \label{esomcc}
  \om_1=\pl_1q_\perp\pm\pl_1q\mp\frac{\pl_1^2\pi}{\pl_1\pi}
  +\frac{P\pl_1X}{\pl_1\pi},
\end{equation}
where $q$ is given by Equation (\ref{espcmc}). The upper and lower
signs in Equations (\ref{epmrup})--(\ref{esomcc}) must be chosen
simultaneously.

{\em The case {\bf D}}. $Y=0$, $\pl_1\pi=0$. 
For this case $\pi=\pi(\tau)$ and $p=0$. For these gauge fixing functions
the constraints become
\begin{align}                                           \label{ecdfid}
  H_0&=e^{2q}\left(V-\frac\rho2m^2X^2\right)
  -\frac1{2\rho}P^2-\frac\rho2\pl_1X^2=0,
\\                                                      \label{ehdcos}
  H_1&=P\pl_1X=0.
\end{align}
We see that the space component of the Lorentz connection does not
enter the constraint equations and remains arbitrary. For two
subcases $P=0$ and $\pl_1X=0$ following from Equation (\ref{ehdcos})
the constraint (\ref{ecdfid}) may be solved with respect to $q$
\begin{align}                                           \label{esodqp}
  2q=&\ln\frac{\rho\pl_1X^2}{2(V-\frac\rho2m^2X^2)}, & P&=0,
\\                                                      \label{esodqx}
  2q=&\ln\frac{P^2}{2\rho(V-\frac\rho2m^2X^2)}, & \pl_1X&=0.
\end{align}
\section{Time components of the geometric variables \label{sreeqz}}
In the previous section we solved the constraints for $q$, $p_\perp$,
and $\om_1$ with $p$, $q_\perp$, and $\pi$ defined by the gauge fixing
function (\ref{esegac}). Making the inverse canonical transformation
(\ref{eolnep}) and (\ref{ecotra}) one immediately gets a general solution
to the constraints in terms of the original variables
\begin{align}                                              \label{ezwplu}
  e_1{}^\pm&=\pm\frac1{\sqrt2}e^{q\pm q_\perp},
\\                                                         \label{emoplu}
  p_\pm&=\frac1{\sqrt2}e^{-q\mp q_\perp}(\pl_1\pi\pm p),
\\                                                         \label{elorcs}
  \om_1&=\pl_1q_\perp-\frac1{\pl_1\pi}(\pl_1p-p\pl_1q-P\pl_1X),
\end{align}
The function $q$ is given by (\ref{esolfq}), (\ref{esohoz}), (\ref{espcmc})
or (\ref{esodqp}), (\ref{esodqx}) in terms of the gauge fixing functions
$\pi$, $p$, and $q_\perp$ for Cases {\bf A}, {\bf B}, {\bf C}, or {\bf D},
respectively. We shall keep it as it stands to treat Cases {\bf A, B, C}
simultaneously. In Case {\bf D} the space component of the
Lorentz connection is not defined by the constraints, and the scalar field
is restricted by Equation (\ref{ehdcos}). This case will be treated
separately in Sections \ref{stmatf} and \ref{shomso}.

The next problem is to reconstruct time components of the Lorentz
connection $\om_0$ and the zweibein $e_0{}^a$. To this end one has to
solve the remaining six equations of motion: (\ref{eqmopi}), (\ref{eqmotp})
and time components of (\ref{eqmolc}), (\ref{eqmozw}). Let us write them
for light-cone components to be as close to the matterless case as possible.
The light-cone components of the energy momentum tensor
$T_{\pm\pm}=e^\al{}_\pm e^\bt{}_\pm T_{\al\bt}$ are
\begin{align*}
  T_{++}&=\frac1{4(e_1{}^+)^2}P_-^2=\frac12e^{-2q-2q_\perp}P_-^2,
\\
  T_{+-}&=T_{-+}=\frac12m^2X^2,
\\
  T_{--}&=\frac1{4(e_1{}^-)^2}P_+^2=\frac12e^{-2q+2q_\perp}P_+^2,
\end{align*}
where $P_\pm$ is given by Equation (\ref{edepxs}).
Afterwards the remaining equations become
\begin{align}                                              \label{erefir}
  &\dot\om_1-\pl_1\om_0-(e_0{}^-e_1{}^+-e_0{}^+e_1{}^-)
  \left(\frac12p^ap_aU'+V'-\frac{\rho'}{2g_{11}}P_+P_-\right)=0,
\\                                                         \label{eresec}
  &\dot e_1{}^+-\pl_1e_0{}^+-\om_0e_1{}^++\om_1e_0{}^+
  -(e_0{}^-e_1{}^+-e_0{}^+e_1{}^-)p_-U=0,
\\                                                         \label{erethi}
  &\dot e_1{}^--\pl_1e_0{}^-+\om_0e_1{}^--\om_1e_0{}^-
  -(e_0{}^-e_1{}^+-e_0{}^+e_1{}^-)p_+U=0,
\\                                                         \label{erefor}
  &\dot\pi-e_0{}^+p_++e_0{}^-p_-=0,
\\                                                         \label{erefif}
  &\dot p_++\om_0p_++e_0{}^-\left(\frac12p^ap_aU+V-\frac12\rho m^2X^2\right)
  +e_0{}^+\frac\rho{4(e_1{}^+)^2}P_-^2=0,
\\                                                         \label{eresix}
  &\dot p_--\om_0p_--e_0{}^+\left(\frac12p^ap_aU+V-\frac12\rho m^2X^2\right)
  -e_0{}^-\frac\rho{4(e_1{}^-)^2}P_+^2=0,
\end{align}
The first three equations are linear inhomogeneous differential equations
for $\om_0$ and $e_0{}^\pm$. The last three are purely algebraic in these
variables. All other geometric variables are known functions and scalars
are arbitrary. Equations (\ref{erefir})--(\ref{eresix}) are dependent
between themselves because there are three identities (\ref{edeplo}),
(\ref{edegco}) as the consequence of the invariance of the action under
local Lorentz rotations and general coordinate transformations. Therefore
there is no need to solve all of them.

Let us consider Cases {\bf A} and {\bf B} with $Y\ne0$. In these cases
$p_-\ne0$ and $p_+\ne0$ due to Equation (\ref{emosgp}). First we solve two
algebraic Equations (\ref{erefor})
and (\ref{eresix}) for $e_0{}^-$ and $\om_0$ in terms of $e_0{}^+$
\begin{align}                                              \label{ezwezm}
  e_0{}^-&=\frac1{p_-}(-\dot\pi+e_0{}^+p_+),
\\                                                         \label{ezweoz}
  \om_0&=\frac1{p_-}\left[\dot p_-
  +\frac{\rho\dot\pi}{4(e_1{}^-)^2p_-}P_+^2
  -e_0{}^+\left(\frac12p^ap_aU+V-\frac12\rho m^2X^2
  +\rho\frac{p_+}{4(e_1{}^-)^2p_-}P_+^2\right)\right].
\end{align}
The next step depends on the value of the determinant
\begin{equation}                                           \label{edezec}
  D=\frac{\pl_1\pi-p}{\pl_1\pi+p}P_-^2-\frac{\pl_1\pi+p}{\pl_1\pi-p}P_+^2
\end{equation}
for the system of linear algebraic Equations (\ref{erefor})--(\ref{eresix})
for $e_0{}^\pm$ and $\om_0$. Note that in the absence of matter this
determinant vanishes identically.

If the determinant is nonzero $D\ne0$
then one may solve the remaining algebraic Equation (\ref{erefif}) and
get the solution for all time components of the geometric variables
\begin{align}                                              \label{ezeplz}
  e_0{}^\pm p_\pm&=\frac2{\rho D}\left[-\frac12\dot Y+Y\dot q
  +\dot\pi\left(\frac12YU+Ve^{2q}-\frac12\rho m^2X^2e^{2q}\right)\right]
  -\frac{\dot\pi}D\frac{\pl_1\pi\pm p}{\pl_1\pi\mp p}P_\pm^2,
\\                                                        \nonumber
  \om_0&=\dot q_\perp-\frac1Y(\pl_1\pi\dot p-p\pl_1\dot\pi)
  +\frac\rho{DY}\dot\pi P_-^2P_+^2
\\                                                         \nonumber
  &-\frac4{\rho DY}\left[-\frac12\dot Y+Y\dot q+\dot\pi\left(\frac12YU
  +Ve^{2q}-\frac12\rho m^2X^2e^{2q}\right)\right]
\\                                                         \nonumber
  &\times\left(\frac12YU+Ve^{2q}-\frac12\rho m^2X^2e^{2q}\right)
\\                                                         \label{eloczc}
  &-\frac1{DY}\left(-\frac12\dot Y+Y\dot q\right)
  \left(\frac{\pl_1\pi-p}{\pl_1\pi+p}P_-^2
  +\frac{\pl_1\pi+p}{\pl_1\pi-p}P_+^2\right),
\end{align}
where the $p_\pm$ are given by (\ref{emoplu}).
The remaining differential equations are fulfilled as the consequence
of gauge invariance given by linear dependencies (\ref{edeplo}),
(\ref{edegco}). Thus the whole system of equations of motion is solved
for arbitrary scalar field.

If the determinant is zero $D=0$ then the remaining algebraic equation
yields the equation for an up to now  arbitrary function $A_m(\tau)$
\begin{equation}                                           \label{edefaf}
  -Y\dot T+\frac Y{2(A_m+W_m)}\left(\dot A_m+\dot W_m
  -\frac{\dot\pi}{\pl_1\pi}\pl_1W_m\right)
  +\frac{\rho\dot\pi}2\frac{\pl_1\pi+p}{\pl_1\pi-p}P_+^2=0.
\end{equation}
In the matterless case $T=0$ and for the gauge fixing function $\pi=\pi(\s)$
depending on the space coordinate only, $\dot W_m=0$, and Equation (\ref{edefaf})
reduces to $\dot A_m=0$ and hence $A_m=\const$.

To find the zero component $e_0{}^+$ for $D=0$ one is free to solve one
of the differential equations. We choose it to be Equation (\ref{eresec}).
Substituting (\ref{ezwezm}), (\ref{ezweoz}) and the solution for the space
components of the geometric variables one finally obtains the first order
differential equation
\begin{equation}                                           \label{ezezwe}
  -\pl_1e_0{}^++e_0{}^+F_m+G_m=0,
\end{equation}
where
\begin{align}
  F_m&=\pl_1\left(q_\perp+\frac12\ln\left|\frac{\pl_1\pi-p}{\pl_1\pi+p}
  (A_m+W_m)\right|-\frac12Q+T\right)
\\
  G_m&=\frac1{\sqrt2}e^{q+q_\perp}\left(2\dot q
  -\frac{\pl_1\dot\pi-\dot p}{\pl_1\pi-p}+\dot\pi U
  -\frac{\rho\dot\pi}{(\pl_1\pi-p)^2}P_+^2\right).
\end{align}
A general solution to this equation is given by
\begin{equation}                                          \label{ezwzpc}
  e_0{}^+=e^{\widetilde F_m}
  \left(C(\tau)+\int\! d\s G_me^{-\widetilde F_m}\right),
\end{equation}
where $\widetilde F_m$ is a primitive
\begin{equation}
  \widetilde F_m=\int\! d\s F_m.
\end{equation}
The remaining time components may be found from (\ref{ezwezm}) and
(\ref{ezweoz}). This provides a solution for all geometric variables
for zero determinant (\ref{edezec}).
\section{Effective action for a scalar field           \label{seffsc}}
In this section we write down effective
equations of motion for a scalar field when all geometric variables are
excluded by means of Equations (\ref{eqmopi})--(\ref{eqmozw}). That is,
we solve all equations of motion for the geometric variables in terms
of a scalar field which is assumed to be arbitrary (this is already
done in previous sections) and substitute this solution back into the
equation for a scalar field. We are able to do this in the Hamiltonian
form. First, we rewrite Equation (\ref{eqmozx}) for a scalar in the
Hamiltonian form
\begin{align}                                           \label{ehamxf}
  \dot X&=\frac e{\rho g_{11}}P+\frac{g_{01}}{g_{11}}\pl_1X,
\\                                                      \label{ehammx}
  \dot P&=\pl_1\left(\frac{e\rho}{g_{11}}\pl_1X\right)
  +\pl_1\left(\frac{g_{01}}{g_{11}}P\right)+e\rho m^2X.
\end{align}

From now on for simplicity we shall use the gauge
\begin{equation}                                        \label{egacon}
  \pi=\pi(\s),~~~~p=0,~~~~q_\perp=0,
\end{equation}
assuming $\pl_1\pi\ne0$. This gauge choice corresponds to Case
{\bf A} of a general situation. Then $Y=\pl_1\pi^2$.
In this gauge expressions for nonlocal quantities
(\ref{edefze})--(\ref{esoqex}) become
\begin{align}                                           \label{ebpaga}
  2T(\tau,\s)&=-\int_\s^\infty\frac{d\s'}{\pl_1\pi}
  \left(\frac1{\rho}P^2+\rho\pl_1X^2\right),
\\                                                      \label{edwfig}
  W_m(\tau,\s)&=\int_0^\s\! d\s'\pl_1\pi
  \left(V-\frac\rho2m^2X^2\right)e^{-Q+2T},
\\                                                      \label{espcme}
  e^{2q}&=\frac{\pl_1\pi^2}{2(A_m+W_m)}e^{-Q+2T}.
\end{align}
The solution (\ref{ezwplu})--(\ref{elorcs}) for the space components
of geometric variables simplifies
\begin{align}                                           \label{ediagg}
  e_1{}^\pm&=\pm\frac{e^q}{\sqrt2},
\\                                                      \label{emosfg}
  p_\pm&=\frac{\pl_1\pi e^{-q}}{\sqrt2},
\\                                                      \label{elorcf}
  \om_1&=\frac1{\pl_1\pi}P\pl_1X.
\end{align}
Remember that we are considering the domain of the space-time where
$A_m+W_m>0$ which in the matterless case corresponds to the domain
outside a horizon of the Schwarzschild black hole.

Differentiating Equation (\ref{edwfig}) with respect to $\s$ and
using (\ref{espcme}) one easily sees that the function $A_m+W_m$
satisfies the following differential equation
\begin{equation}                                        \label{edifaw}
  \pl_1(A_m+W_m)=\frac{2e^{2q}}{\pl_1\pi}
  \left(V-\frac12\rho m^2X^2\right)(A_m+W_m)
\end{equation}
which will be used in the following section.

Reconstruction of time components of the geometric variables depends
on the value of the determinant (\ref{edezec}) which is considerably
simplified in the gauge (\ref{egacon})
\begin{equation}                                        \label{edetfg}
  D=-\frac4\rho P\pl_1X.
\end{equation}
Roughly speaking in the considered gauge it coincides with the momentum
density of a scalar field. We consider the cases $D\ne0$ and $D=0$
separately. Note that the case of zero determinant coincides with the
constraint $H_1=0$ Equation (\ref{ehdcos}) in Case {\bf D}.
\subsection{Moving scalar field $D\ne0$                \label{smomat}}
If the determinant (\ref{edetfg}) is nonzero $D\ne0$ then the scalar
field moves. In this case to reconstruct the time components of
geometric variables one has to use Equations (\ref{ezeplz})--(\ref{eloczc}).
Straightforward calculations yield
\begin{align}                                        \label{etizwf}
  e_0{}^\pm&=-\frac{\pl_1\pi\dot qe^q}{\sqrt2 P\pl_1X},
\\                                                   \label{etclog}
  \om_0&=\frac{\dot q}{P\pl_1X}\left(\frac12\pl_1\pi^2U+Ve^{2q}
  -\frac\rho2m^2X^2e^{2q}+\frac1{2\rho}P^2+\frac\rho2\pl_1X^2\right).
\end{align}
The relation $e_0{}^+=e_0{}^-$ follows from Equation (\ref{ezwezm})
because $p_+=p_-$ and $\dot\pi=0$ in the chosen gauge (\ref{egacon}).
From (\ref{ediagg}) and (\ref{etizwf}) one gets expressions for the
determinant of the zweibein
\begin{equation}                                        \label{ezwede}
  e=-\frac{\pl_1\pi\dot qe^{2q}}{P\pl_1X}
\end{equation}
and the metric
\begin{equation}                                        \label{emetfg}
  ds^2=e^{2q}\left(\frac{\pl_1\pi^2\dot q^2}{(P\pl_1X)^2}d\tau^2-d\s^2\right).
\end{equation}

Substitution of the geometric variables into the equations of motion
for a scalar field (\ref{ehamxf}), (\ref{ehammx}) yields the effective
equations
\begin{align}                                           \label{edoxef}
  \dot X&=\frac{\pl_1\pi\dot q}{\rho P\pl_1X}P,
\\                                                      \label{edopgf}
  \dot P&=\pl_1\left(\frac{\rho\pl_1\pi\dot q}{P\pl_1X}\pl_1X\right)
  -\rho m^2\frac{\pl_1\pi\dot qe^{2q}}{P\pl_1X}X.
\end{align}
They may be rewritten in an equivalent and more suitable form, which does
not contain the time derivative $\dot q$. Indeed, the function
$\pl_1\pi^2\dot q/P\pl_1X$ satisfies the following ordinary differential
equation
\begin{equation}                                        \label{edotqp}
  \pl_1\left(\frac{\pl_1\pi^2\dot q}{P\pl_1 X}\right)
  =2\frac{e^{2q}}{\pl_1\pi}\left(V-\frac\rho2m^2X^2\right)
  \frac{\pl_1\pi^2\dot q}{P\pl_1X}.
\end{equation}
This is a crucial observation, and therefore we sketch the proof.
Straightforward calculation of the left hand side of (\ref{edotqp}) and
elimination of the time derivatives $\dot X$ and $\dot P$ using
equations of motion (\ref{edoxef}), (\ref{edopgf}) yields
$$
  \pl_1\left(\frac{\pl_1\pi^2\dot q}{P\pl_1X}\right)
  =2\pl_1\left(\frac{\pl_1\pi^2\dot q}{P\pl_1X}\right)
  -2\pl_1\pi e^{2q}\left(V-\frac\rho2m^2X^2\right)
  \frac{\dot q}{P\pl_1X}.
$$
This is equivalent to Equation (\ref{edotqp}). The same equation is
satisfied by the function $A_m+W_m$ (\ref{edifaw}). Therefore
they may differ only by an arbitrary function of time which is absorbed
by a redefinition of $A_m$. Hence without loss of generality we set
\begin{equation}                                        \label{eqdoaw}
  \frac{\pl_1\pi^2\dot q}{P\pl_1 X}=\frac1{\sqrt\kappa}(A_m+W_m).
\end{equation}
Here we introduced the coupling constant $\kappa$ for dimensional reasons.
This relation is valid only when the effective equations of motion for
scalars are satisfied. Afterwards Equations (\ref{edoxef}), (\ref{edopgf})
and the metric (\ref{emetfg}) are rewritten in an equivalent form
\begin{align}                                           \label{edxeff}
  \sqrt\kappa\dot X&=\frac1{\rho\pl_1\pi}(A_m+W_m)P,
\\                                                      \label{edopef}
  \sqrt\kappa\dot P&=\pl_1\left[\frac\rho{\pl_1\pi}(A_m+W_m)\pl_1X\right]
  -\frac12\pl_1\pi e^{-Q+2T}\rho m^2X,
\\                                                      \label{elieaw}
  ds^2&=\frac12e^{-Q+2T}
  \left[\frac1\kappa(A_m+W_m)d\tau^2-\frac{d\pi^2}{A_m+W_m}\right].
\end{align}
For the matterless case $T=0$, $A_m=A=\const$, $W_m=W$, and the last
expression for the line element coincides with (\ref{emaled}).
For spherically reduced gravity in the matterless case it yields
the Schwarzschild solution (see Section \ref{sredgr}).

Using relation (\ref{eqdoaw}) the solution for the time components
(\ref{etizwf}), (\ref{etclog}) becomes
\begin{align}                                        \label{etizws}
  e_0{}^\pm&=-\frac1{\sqrt{2\kappa}}\frac{A_m+W_m}{\pl_1\pi}e^q,
\\                                                   \label{etclos}
  \om_0&=\frac{A_m+W_m}{\sqrt\kappa\pl_1\pi^2}
  \left(\frac12\pl_1\pi^2U+Ve^{2q}
  -\frac\rho2m^2X^2e^{2q}+\frac1{2\rho}P^2+\frac\rho2\pl_1X^2\right).
\end{align}
The important point is that the effective equations of motion for a
scalar field (\ref{edxeff}), (\ref{edopef}) do not
contain any metric components and are written entirely in terms of
scalar fields $X$, $P$, and the gauge fixing function $\pi(\s)$. All
geometric variables are expressed in terms of the scalar field only by
Equations (\ref{ediagg})--(\ref{elorcf}) and (\ref{etizws}), (\ref{etclos}).
In particular, the two-dimensional scalar curvature and torsion squared
term are
\begin{align}                                           \label{escexe}
  R&=e^{-2q}\left(-\frac{\rho'}\rho P^2+\rho'\pl_1X^2+U'\pl_1\pi^2\right)
  -2V'+\rho'm^2X^2,
\\                                                      \label{etoexe}
  T^*_aT^{*a}&=4e^{-2q}\pl_1\pi^2U^2.
\end{align}

In other words we have proved the following theorem.
\begin{Theorem}
Any solution of the matter-gravity equations of motion
(\ref{eqmopi})--(\ref{eqmozx}) in the gauge (\ref{egacon}) has the form
(\ref{ediagg})--(\ref{elorcf}), (\ref{etizws}), (\ref{etclos}) where
the scalar field satisfies the effective equations of motion
(\ref{edxeff}), (\ref{edopef}) under the condition $P\pl_1X=0$.
Inversely, for arbitrary solution of the effective equations of motion
for a scalar field with $P\pl_1X\ne0$ the zweibein and Lorentz connection
constructed by Equations (\ref{ediagg})--(\ref{elorcf}), (\ref{etizws}),
(\ref{etclos}) satisfies the original equations of motion
(\ref{eqmopi})--(\ref{eqmozx}).
\end{Theorem}

Note that the Lorentz connection does not enter the effective
equations for scalars (\ref{edxeff}), (\ref{edopef}) as it should be
because a scalar field does not feel Lorentz rotations.

There arises a natural question whether the effective action leading
to the effective equations of motion exists or not. The answer is positive
\begin{equation}                                        \label{effact}
  S_{\rm eff}=\int\!d\tau\left(\int_0^\infty\!d\s P\dot X
  -H_{\rm eff}\right),
\end{equation}
where the effective Hamiltonian has the form
\begin{equation}                                        \label{effham}
  H_{\rm eff}=\frac1{\sqrt\kappa}\int_0^\infty\!d\s\left[
  \frac{A_m}{\pl_1\pi}\left(\frac1{2\rho}P^2+\frac\rho2\pl_1X^2\right)
  -\frac12\pl_1\pi\left(V-\frac12\rho m^2X^2\right)e^{-Q+2T}\right].
\end{equation}
One may check by straightforward calculations that variation of this
effective action yields the effective equations of motion for
scalars (\ref{edxeff}), (\ref{edopef}). The limits of integration in
(\ref{effham}), (\ref{ebpaga}), and (\ref{edwfig}) are adapted for
spherically reduced gravity considered in Section \ref{sredgr}. In
other models they may be changed. The unusual feature of
the effective action is its nonlocality due to the term with $T$.

The intriguing point is the following. The Hamiltonian (\ref{ehamsd})
of the original system is the linear combination of the constraints.
If one solved the equations of motion for the geometric variables and
substituted the solution back into the Hamiltonian one would get zero
because the constraints constitute part of the equations of motion.
The existence of the effective action (\ref{effham}) is a very
important observation because the value of $H_{\rm eff}$ is
conserved for any trajectory is the phase space. It may be helpful
for the analysis of the equations of motion and quantization of the
model.
\subsection{An example                                    \label{sexamp}}
The effective action for a scalar field obtained in previous sections
is produced by the boundary term which must be added to the original
action (\ref{efoact}). To clarify its appearance we consider a simple
example in this section. Notations here have nothing to do
with the rest of the paper.

Let the model be described by two sets of canonically conjugate variables
$q,p$ and $q^*,p^*$ depending on $\tau$ and $\s$.
Consider the action written in the Hamiltonian form
\begin{equation}                                        \label{eacexa}
  S=\int_{-\infty}^\infty\!\!d\tau\int_0^\infty\!\!d\s
  (p\dot q+p^*\dot q^*-\lm G),
\end{equation}
where the Hamiltonian is given by a single constraint
$$
  H=\int_0^\infty\!\!d\s \lm G,~~~~~~G=-\pl_1q+H^*(q^*,p^*).
$$
Here $H^*(q^*,p^*)\ge0$ is some nonnegative function which does not contain
space derivatives of its arguments, and $\lm$ is a Lagrange multiplier.
For definiteness we specify the limits of integration.

One may pose two variational problems for the action (\ref{eacexa}).
The most common one assumes that variations of all fields are compactly
supported functions. In this case one can freely integrate by parts
and drop all boundary terms. In the second variational problem
variations of the fields are not assumed vanishing at the limits of
integration.
For this problem the action is sensitive to addition of boundary terms,
and besides the equations of motion one obtains boundary conditions on
the fields. For example, this problem is considered in the open bosonic
string theory where the variational problem produces equations of motion
and Neumann boundary conditions.

Let us consider the first variational problem for the action (\ref{eacexa}).
Then one obtains only equations of motion
\begin{align}                                           \label{eqdotf}
  \dot q&=0,
\\                                                      \label{epdotf}
  \dot p&=-\pl_1\lm,
\\                                                      \label{eqsdot}
  \dot q^*&=\lm\frac{\pl H^*}{\pl p^*},
\\                                                      \label{epsdot}
  \dot p^*&=-\lm\frac{\pl H^*}{\pl q^*},
\\                                                      \label{ephico}
  G&=-\pl_1q+H^*(q^*,p^*)=0.
\end{align}
Equation (\ref{ephico}) is a constraint on the canonical variables,
and it is obviously of the first class
$$
  \lbrace G,G'\rbrace=0,
$$
where $G'=G(\tau,\s')$. Therefore $\lbrace G,H\rbrace=0$, and there are
no other constraints.
To eliminate the unphysical (nonpropagating) degree of freedom one has to
impose one gauge condition. In the canonical gauge fixing approach
\cite{GitTyu90,HenTei92} it must form a system of second class constraints
together with $G$. We choose it to be
$$
  F=p-p(\s)=0,
$$
where $p(\s)$ is a given function of space coordinate only.

We see that a pair of canonical variables $q,p$ do not describe a physical
degree of freedom and may be eliminated from the model explicitly.
To this end we solve the equations of motion and the constraint in the
chosen gauge. First we solve the constraint (\ref{ephico})
\begin{equation}                                        \label{esocoq}
  q=\int_0^\s\!\!d\s'H^*(q^*,p^*)+q_0(\tau),
\end{equation}
where $q_0(\tau)$ is an arbitrary function of time. To be consistent
with the equation of motion (\ref{eqdotf}) the following equation must
hold $\dot q_0=0$. Thus $q_0=\const$, and is insignificant. The Lagrange
multiplier is defined by Equation (\ref{epdotf}) with $\dot p=0$
which has a general solution
\begin{equation}                                        \label{elames}
  \lm=\lm_0(\tau),
\end{equation}
where $\lm_0(\tau)$ is an arbitrary function to be fixed, for example,
by a boundary condition.

We see that the model describes a single physical degree of freedom
$q^*,p^*$ with the effective Hamiltonian $\lm_0 H^*(q^*,p^*)$. There are
two important observations. First, the effective Hamiltonian has
the usual form and was obtained by local elimination of the
unphysical degree of freedom. It means that the effective Hamiltonian
does not depend on whether the space is closed (a circle) or open (line).
Second, the effective Hamiltonian cannot be obtained from the action
(\ref{eacexa}) by going to the constraint surface in the fixed gauge
because
$$
  S\Big|_{F=0,G=0}=\int\!\! d\tau d\s\,p^*\dot q^*.
$$

To resolve the problem let us consider the second variational problem.
We assume that variations of the physical degree of freedom $q^*$, $p^*$
and the Lagrange multiplier are compactly supported. We are free to do this
because the physical degree of freedom is not restricted by the constraint.
At the same time the variation of the unphysical degree of freedom
$\dl^* q(\tau,\s)$ caused by the variations $\dl q^*$ and $\dl p^*$
has the form
$$
  \dl^*q=\int_0^\s\!\!d\s'\left(\frac{\pl H^*}{\pl q^*}\dl q^*
  +\frac{\pl H^*}{\pl p^*}\dl p^*\right).
$$
We see that this variation cannot be compactly supported in space as
the consequence of the constraint equation. Therefore we are not allowed
to drop boundary terms in space containing $q$. At the same time the
constraint does not restrict variations of $q$ at time boundaries, and
we assume that $\dl^*q(\pm\infty,\s)=0$. Consider the new action
\begin{equation}                                        \label{enweac}
  S_{\rm ph}=S-\int\!\!d\tau\lm q\Big|_{\s=0}^\infty
\end{equation}
differing from the old one by the boundary term. This boundary term
is chosen in such a way as to compensate the boundary contribution
to the variation of $S$ with respect to $q$. Therefore the variation
of the action (\ref{enweac}) yields the same equations of motion
(\ref{epdotf})--(\ref{ephico}), and no boundary conditions arise even
for nonvanishing variations $\dl q$. The effective Hamiltonian is now
obtained by going to the constraint surface
\begin{equation}                                        \label{efface}
  S_{\rm ph}\Big|_{F=0,G=0}=\int\!\! d\tau\int_0^\infty\!\!d\s
  \left(p^*\dot q^*-\lm(\tau,\infty) H^*(q^*,p^*)\right).
\end{equation}
The boundary term in the upper limit is given by the integral over the
whole space (\ref{esocoq}) and hence produce the effective Hamiltonian
with a well defined Hamiltonian density. The reason for this is that the
constraint is a
differential equation which does not admit compactly supported solutions.

The above consideration is valid also for a compact space. Let
$\s\in(0,2\pi)$. We assume that the physical degree of freedom is
described by smooth functions $q^*$ and $p^*$ given on a circle.
The unphysical degree of freedom and its variation cannot be smooth
functions on a circle because of the constraint (\ref{esocoq}), and
a cut with the appropriate boundary condition must be done.
Therefore for a circle one simply has to change the limits of
integration over $\s$ in (\ref{efface}).

In a more general gauge
$$
  F=p-p(\tau,\s)=0,
$$
where a function $p(\tau,\s)$ may depend on time, the effective action
acquires an additional contribution due to the term $p\dot q$. One can
easily check that this contribution does not change the final answer. Indeed,
$$
  \triangle S\Big|_{F=0,G=0}=\int\!\!d\tau d\s\,p\dot q
  =\int\!\!d\tau d\s\,(-\dot p q).
$$
Using Equation (\ref{epdotf}) which defines $\lm$ and integrating by
parts we get
$$
  \triangle S\Big|_{F=0,G=0}=\int\!\!d\tau d\s(-\lm\pl_1q)
  +\int\!\!d\tau\,\lm q\Big|_{\s=0}^\infty.
$$
The first term now reproduces the effective Hamiltonian $\lm H^*$ and
the second one is cancelled by the boundary term (\ref{enweac}).
We see that the time dependent gauge also produces correctly the effective
Hamiltonian. The considered example shows that this procedure is not
the only one, and the origin of the nontrivial Hamiltonian for the
physical degree of freedom lies in the boundary term. It arises in the
analysis of the constraint equation which does not admit compactly
supported solutions.
\subsection{The effective action as a boundary term    \label{sefabo}}
The example considered in the previous section describes correctly a
general situation in gravity models. In our case the constraints $G=0$
and $H_1=0$ in Section \ref{scosol} were solved algebraically with
respect to $p_\perp$ and $\om_1$. The constraint $H_0=0$ is a differential
equation with respect to $q$, and its solution (\ref{esolfq}) cannot be
compactly supported. The Hamiltonian (\ref{ehamsd}) is equal to
$$
  H=\int\!\!d\s\left[\left(\om_0+\frac{e-g_{01}}{g_{11}}\om_1\right)G
  -\frac e{g_{11}}H_0+\frac{g_{01}}{g_{11}}H_1\right].
$$
The only term which contains spatial derivative of $q$ is the second one.
Therefore to compensate the boundary contribution due to the variation
of $q$ we add the boundary term to the action (\ref{efoact})
\begin{equation}                                        \label{eboure}
  \triangle S=-\int\!\!d\tau\frac e{g_{11}}\pl_1\pi q\Big|_{\s=0}^\infty.
\end{equation}
Expressing everything in terms of the scalar field and gauge fixing
functions we get
$$
  \frac e{g_{11}}\pl_1\pi=\frac1{\sqrt{\kappa}}(A_m+W_m).
$$
Afterwards the boundary terms are easily calculated
\begin{align*}
  -\frac1{\sqrt\kappa}(A_m+W_m)q\Big|_{\s=\infty}&=
  \frac1{2\sqrt\kappa}\int_0^\infty\!\!d\s\,\pl_1\pi
  \left(V-\frac12\rho m^2X^2\right)e^{-Q+2T},
\\
  -\frac1{\sqrt\kappa}(A_m+W_m)q\Big|_{\s=0}&=
  \frac1{2\sqrt\kappa}A_m\int_0^\infty\frac{d\s}{\pl_1\pi}
  \left(\frac1\rho P^2+\rho\pl_1X^2\right).
\end{align*}
They produce exactly the effective action (\ref{effact}).

The procedure of obtaining the effective action was noted in
\cite{Unruh76} in the case of spherically reduced gravity.
\subsection{Static scalar field $P=0$                  \label{stmatf}}
The equation $D=0$ for zero determinant (\ref{edetfg}) has two solutions
$P=0$ and $\pl_1X=0$ corresponding
to static and homogeneous distribution of a scalar matter. In this and
the following sections we consider these cases, respectively.
They yield another classes of solutions to the whole system of the
equations of motion.

For $D=0$ Equation (\ref{edefaf}) in the gauge (\ref{egacon})
reduces to
\begin{equation}                                        \label{eqarfa}
  2\dot T=\frac{\dot A_m+\dot W_m}{A_m+W_m}~~\Leftrightarrow~~\dot q=0,
\end{equation}
and restricts an arbitrary function $A_m(\tau)$.
Equation (\ref{ezezwe}) for the time component of the zweibein $e_0{}^+$
takes a simple form
\begin{equation}                                        \label{etipzw}
  -\pl_1e_0{}^++e_0{}^+F_m=0,
\end{equation}
where
$$
  F_m=\pl_1\left[\frac12\ln(A_m+W_m)-\frac12Q+T\right].
$$
It has a general solution (\ref{etizws}) up to a nonzero factor depending
on $\tau$ which may be absorbed by a redefinition of time coordinate.
Expressions for the Lorentz connection $\om_0$ and space components of
geometric variables remain the same as before (\ref{etclos}),
(\ref{ediagg})--(\ref{emosfg}) with $\om_1=0$. The corresponding metric
takes the form (\ref{elieaw}) as in the previous case of moving scalars.

Then the effective equations of motion for a scalar field take
precisely the form (\ref{edxeff}), (\ref{edopef}) as in the previous
case. Equation (\ref{edxeff}) together with the restriction $P=0$
yields $\dot X=0$. Hence the solution is static $X=X(\s)$.
We see that the functions $T(\s)$ and $W_m(\s)$ defined by (\ref{ebpaga})
and (\ref{edwfig}) depend on the space coordinate only. Therefore Equation
(\ref{eqarfa}) yields $A_m=\const$.

Afterwards Equation (\ref{edopef}) defines the static distribution
of a scalar field
\begin{equation}                                        \label{estdsc}
  \pl_1\left[\frac\rho{\pl_1\pi}(A_m+W_m)\pl_1X\right]
  -\frac12\pl_1\pi e^{-Q+2T}\rho m^2X=0.
\end{equation}
It provides a general solution to the whole system of equations of
motion in the static case. In the massless case $m=0$ this equation
simplifies to
\begin{equation}                                        \label{emzest}
  \frac{\rho}{\pl_1\pi}(A_m+W_m)\pl_1X=\const.
\end{equation}
It may be rewritten in a local form. Differentiating it two times with
respect to $\s$ and assuming $\pl_1X\ne0$ one obtains the equation
\begin{equation}                                        \label{efeqst}
  \frac{\pl_1^2 Z}{\pl_1 Z}-\frac{\pl_1\pi}{\rho Z^2}
  -\frac{\pl_1^2\pi}{\pl_1\pi}+U\pl_1\pi-\frac{\pl_1V}V=0,
\end{equation}
where
$$
  Z(\s)=\frac{\pl_1\pi}{\rho\pl_1X}.
$$
This equation will be solved in Section \ref{stsosp} for spherically
reduced gravity.
\subsection{Homogeneous solution $\pl_1X=0$            \label{shomso}}
For the homogeneous solution the scalar field depends only on time
coordinate $X=X(\tau)$. Expressions for all geometric quantities
$e_1{}^\pm$, $\om_1$ and $e_0{}^\pm$, $\om_0$ are the same as in the
previous case  (\ref{ediagg}), (\ref{elorcf}) and (\ref{etizws}),
(\ref{etclos}). The effective equations of motion reduce to
\begin{align}                                           \label{ehoscf}
  \sqrt\kappa\dot X&=\frac1{\rho\pl_1\pi}(A_m+W_m)P,
\\                                                      \label{ehoscm}
  \sqrt\kappa\dot P&=-\frac12\pl_1\pi e^{-Q+2T}\rho m^2X.
\end{align}
This system of equations defines the homogeneous solution $X=X(\tau)$ and
$P=P(\tau,\s)$. In general the momenta may depend on both coordinates.

The solution simplifies for zero mass $m=0$. Then Equation (\ref{ehoscm})
yields $P=P(\s)$, and hence nonlocal quantities (\ref{ebpaga}) and
(\ref{edwfig}) depend only on space coordinate $T=T(\s)$ and $W_m=W_m(\s)$.
In this case Equation (\ref{eqarfa}) yields a solution $A_m=\const$ as
in the static case. We see that the right hand side of Equation
(\ref{ehoscf}) depends on the space coordinate only whereas the left hand
side depends only on time. This is possible only for linear
dependence on time
\begin{equation}                                        \label{ehosox}
  X=a\tau+b,~~~~~~a,b=\const.
\end{equation}
Afterwards we obtain the equation for $P=P(\s)$
\begin{equation}                                        \label{ehoemo}
  \frac1{\rho\pl_1\pi}(A_m+W_m)P=a,
\end{equation}
which is similar to Equation (\ref{emzest}) for the static case.
This nonlocal equation may be written in a local form after
differentiating it two times. Introducing a new variable for $P\ne0$
$$
  Z(\s)=\frac{\rho\pl_1\pi}P
$$
one gets
\begin{equation}                                        \label{eqyhom}
  \frac{\pl_1^2Z}{\pl_1Z}-\frac{\rho\pl_1\pi}{Z^2}
  -\frac{\pl_1^2\pi}{\pl_1\pi}+U\pl_1\pi-\frac{\pl_1 V}V=0
\end{equation}
which differs from (\ref{efeqst}) in the second term only.

We see that in both degenerate cases $P=0$ or $\pl_1X=0$ one has to
solve the same system of the effective equations of motion
(\ref{edxeff}), (\ref{edopef}), and the metric has the same form
(\ref{elieaw}) as in a general case. The only difference is that
$A_m$ must be constant for static and homogeneous solutions for a
massless scalar field.
\section{Spherically reduced gravity                   \label{sredgr}}
Let us consider spherically reduced general relativity which is of
great importance for our understanding of a black hole formation. We
start with the Hilbert--Einstein action proportional to the
four-dimensional scalar curvature $R^{(4)}$ minimally coupled to a
massive scalar field. The four-dimensional Lagrangian is
\begin{equation}                                        \label{ehieis}
  L=\kappa eR^{(4)}-2\Lm e+\frac12e(g^{ij}\pl_iX\pl_jX-m^2X^2),
  ~~~~~e=\sqrt{|\det g_{ij}|},
\end{equation}
where $i,j=0,1,2,3$, $\kappa>0$, and the $\Lm$ are gravitational and
cosmological constants. Assuming a spherical symmetry for a metric
\begin{equation}                                        \label{espsym}
  ds^2=g_{\al\bt}dx^\al dx^\bt+\frac1{2\kappa}\pi d\Om
\end{equation}
and for a scalar field
\begin{equation}                                        \label{escfis}
  X=X(x^\al),~~~~\al=0,1,
\end{equation}
where
$$
  d\Om=d\theta^2+\sin^2\theta d\f^2
$$
is the usual metric on a unit sphere, the Lagrangian (\ref{ehieis}) is
reduced to
\begin{equation}                                        \label{espreh}
  L=e\left[-\frac12\pi\widetilde R+\frac{\pl\pi^2}{4\pi}
  +2\kappa+\frac{\Lm\pi}\kappa-\frac\pi{4\kappa}
  (\pl X^2-m^2X^2)\right].
\end{equation}
Here $\widetilde R$ is the two-dimensional scalar curvature for a
two-dimensional metric $g_{\al\bt}$. We introduced a gravitational
constant in the spherically symmetric ansatz for the metric (\ref{espsym})
in order for the dilaton field to be dimensionless.
Thus a minimally coupled scalar field in four dimensions after spherical
reduction yields the two-dimensional scalar field nonminimally coupled to
two-dimensional dilaton gravity. The Lagrangian (\ref{espreh}) coincides
with (\ref{edilmo}) for the functions $U$ and $V$ given by (\ref{echfsr}).
We restrict ourselves to negative values of the dilaton field $\pi<0$ to
provide the four-dimensional metric with the signature $(+---)$.

The spherical reduction of the four-dimensional action (\ref{ehieis}) to
the two-dimensional one (\ref{espreh}) was done on the level of the actions.
In general this procedure is not equivalent to a reduction at the level of
the equations of motion. In the case of spherically reduced gravity both
reductions are equivalent \cite{ThIsHa84}. That is, the substitution of
(\ref{espsym}) and (\ref{escfis}) in the four-dimensional equations of
motion yields the system of equations which is equivalent to the equations
of motion following directly from the two-dimensional Lagrangian
(\ref{espreh}).

In this section we consider the case of a massless scalar field $m=0$ and
zero cosmological constant $\Lm=0$ which has attracted much interest in
the last years due to the discovery of critical phenomena in gravitational
collapse \cite{Choptu93} (for a recent review see \cite{Gundla00}). Note
also that this matter model may be interpreted in terms of "superstiff"
perfect liquid with the energy density$=$pressure equation of state
\cite{BelHal72}. For comparison we choose the Schwarzschild like coordinates
\begin{equation}                                        \label{egaupi}
  \pi=-2\kappa\s^2,~~~~\s\in(0,\infty),
\end{equation}
with the metric (\ref{espsym}) is of the form
\begin{equation}                                        \label{efodim}
  ds^2=g_{\al\bt}dx^\al dx^\bt-\s^2d\Om.
\end{equation}
For this gauge choice functions defining the two-dimensional
gravity model (\ref{echfsr}) become
\begin{equation}                                        \label{espspm}
  U=-\frac1{4\kappa\s^2},~~~~V=-2\kappa,~~~~\rho=-\s^2.
\end{equation}
For spherically reduced gravity we choose the primitive (\ref{eqwfun})
to be $Q=\ln|\pi|/2$ which in the gauge (\ref{egaupi}) yields
$$
  e^{-Q}=\frac1{\sqrt{2\kappa}\,\s}.
$$

To simplify the following analysis we introduce dimensionless coordinates
and a scalar field
\begin{equation}                                        \label{edimco}
\begin{aligned}
  \s&=\frac{r}{\sqrt\kappa},
\\
  X&=2\sqrt\kappa\widetilde X,
\end{aligned}
\qquad
\begin{aligned}
  \tau&=\frac t{\sqrt{2\kappa}},
\\
  P&=2\widetilde P.
\end{aligned}
\end{equation}
Next we redefine the nonlocal quantity
\begin{equation}                                        \label{eredwm}
  A_m+W_m=4\sqrt2\,\kappa(\widetilde A_m+\widetilde W_m).
\end{equation}
Dropping tilde signs we arrive at the effective equations of motion
\begin{align}                                           \label{escspx}
  \dot X&=\frac1{r^3}(A_m+W_m)P,
\\                                                      \label{escspp}
  \dot P&=\pl_1\left[(A_m+W_m)r\pl_1X\right],
\\ \intertext{where dots denote derivatives with respect to time $t$,
$\pl_1=\pl_r$, and}
  2T(t,r)&=-\int_r^\infty\frac{d\s}\s                   \label{etspgr}
  \left(\frac1{\s^2}P^2+\s^2\pl_1X^2\right)\le0,
\\                                                      \label{exdilw}
  W_m(t,r)&=\int_0^rd\s e^{2T}\ge0.
\end{align}
Both functions are monotonically increasing functions of $r$ with the
boundary values $T(t,\infty)=0$ and $W_m(t,0)=0$. The following bound
holds $W_m(t,r)\le r$ because $T$ is nonpositive.
This is a full set of the effective equations of motion for spherically
reduced gravity coupled to a massless scalar field. Here $A_m=A_m(t)$
is an arbitrary function of time to be defined by a boundary condition.
The effective Hamiltonian (\ref{effham}) leading to the effective
equations of motion (\ref{escspx}), (\ref{escspp}) has the form
\begin{equation}                                        \label{effacs}
\begin{split}
  H_{\rm eff}&=\frac12\int_0^\infty\!dr\left[\frac{A_m}r\left(
  \frac1{r^2}P^2+r^2\pl_1X^2\right)+1-e^{2T}\right]
\\
  &=-A_m(t)T(t,0)+\frac12\left(\int_0^\infty\!dr-W_m(t,\infty)\right).
\end{split}
\end{equation}
Here we inserted the unity in the integrand which does not alter the
equations of motion for normalization. The reason for this will
become clear later from the comparison with the free spherical
waves in flat Minkowskian space-time. The expression for the effective
Hamiltonian may be rewritten in the other form. We integrate by parts
$$
  \int_0^\infty\!dr(1-e^{2T})=r(1-e^{2T})\Big|_0^\infty
  +\int_0^\infty\!dre^{2T}\left(\frac1{r^2}P^2+r^2\pl_1X^2\right).
$$
The first term in the right hand side may be dropped if the integral
(\ref{etspgr}) converges at the infinity. Then the effective Hamiltonian
is
\begin{equation}                                        \label{effhan}
  H_{\rm eff}=\frac12\int_0^\infty\!dr
  \left(e^{2T}+\frac{A_m}r\right)
  \left(\frac1{r^2}P^2+r^2\pl_1X^2\right).
\end{equation}
The Hamiltonian density is clearly positive definite if
\begin{equation}                                        \label{eresta}
  e^{2T}+\frac{A_m}r>0.
\end{equation}

Let us make a remark concerning the positive energy theorem in
asymptotically flat space-time in general relativity (for review,
see \cite{Faddee82,GitTyu90}). The theorem states that the total
energy defined via the surface integral is positive under the
reasonable assumptions on the matter energy-momentum tensor.
The assumption of asymptotic flatness was essential because for
closed universes the surface integrals were assumed to vanish.
In the present paper we showed that this assumption is not valid
because unphysical degrees of freedom cannot be compactly supported
functions as a solution of the constraint equations. In our case we
have the positive definite Hamiltonian density which is the same for
open and closed Universes, and expression (\ref{effhan}) proves
the positive energy theorem for spherically symmetric solutions
under the restriction (\ref{eresta}).

The effective Hamiltonian explicitly depends on time through the
function $A_m(t)$. Therefore the energy is not conserved
$$
  \frac{dE}{dt}=\frac{\dot A_m}2\int_0^\infty\!\frac{dr}r
  \left(\frac1{r^2}P^2+r^2\pl_1 X^2\right).
$$

The effective Hamiltonian (\ref{effacs}) for $A_m=0$ was obtained
in \cite{Unruh76}. The arbitrary function $A_m(\tau)$ arose in a
solution of the constraints which are differential equations with
respect to $\s$ and cannot be eliminated by a reparameterization
of the time coordinate. It may be fixed by a boundary condition on
metric components. The effective equations of motion
(\ref{escspx}), (\ref{escspp}) are similar to the equations of
Section 3 in \cite{Christ86} obtained for the spherically reduced
gravity in the second order formulation assuming regularity at the
center $r=0$. In the present paper the effective equations for
scalars are written in the first order formulation in a general
case and for a wider class of models (\ref{efoact}).

The solution for a metric (\ref{elieaw}) in the spherically reduced
gravity takes the form
\begin{equation}                                        \label{espemd}
  ds^2=\frac1\kappa e^{2T}
  \left(\frac{A_m+W_m}rdt^2-\frac r{A_m+W_m}dr^2\right).
\end{equation}

In general relativity the mass function $M(\tau,\s)$ in the Schwarzschild
like coordinates is usually defined by the equations
$$
  g_{00}=1-\frac{2M}r~~~~~~\text{or}~~~~~~
  -g_{11}=\left(1-\frac{2M}r\right)^{-1}.
$$
For the solution (\ref{espemd}) it is
$$
  2M=r-e^{2T}(A_m+W_m)~~~~~~\text{or}~~~~~~2M=r-e^{-2T}(A_m+W_m).
$$
Since $T(\tau,\infty)=0$, both definitions yield the same expression
for the total mass $M_\infty=M(\tau,\infty)$
$$
  2M_\infty=-A_m+\int_0^\infty\!\!dr(1-e^{2T}).
$$
For $A_m=0$ the total mass equals the total energy
$M_\infty=H_{\rm eff}$ and is conserved in time.

In the case of spherically reduced gravity the effective equations of
motion (\ref{escspx}), (\ref{escspp}) are invariant under the scaling
transformation reflecting the absence of mass (length) parameter in the
model
\begin{align}                                           \nonumber
  t'&=kt, & X'(t',r')&=X(t,r),
\\                                                      \label{escatr}
  r'&=kr, & P'(t',r')&=kP(t,r),
\\                                                      \nonumber
  A^\prime_m(t')&=kA_m(t),  &&
\end{align}
parameterized by a nonzero constant $k=\const$. It means that if
you have some solution to the equations of motion $X(t,r)$,
$P(t,r)$ then the primed functions $X'(t',r')$, $P'(t',r')$
satisfy the same set of equations. Under this transformation the
integrals (\ref{etspgr}), (\ref{exdilw}) transform as
\begin{align*}
  T'(t',r')&=T(t,r),
\\
  W^\prime_m(t',r')&=W_m(t,r).
\end{align*}
\subsection{Small scalar field and asymptotics         \label{smscfi}}
The zero approximation for a small scalar field is the matterless case
$X=0$ and $P=0$ which is obviously a solution to the effective equations
of motion (\ref{escspx}), (\ref{escspp}). Then nonlocal quantities
(\ref{etspgr}), (\ref{exdilw}) become
$$
  T=0,~~~~W_m=r,
$$
Choosing $A_m=-2M$ which must be constant in the matterless case we
immediately get the time-radial part of the Schwarzschild metric
\begin{equation}                                        \label{eshwme}
  ds^2=\frac1\kappa\left[\left(1-\textstyle{\frac{2M}r}\right)dt^2
  -\frac{dr^2}{1-\frac{2M}r}\right].
\end{equation}
Of course for $M=0$ the Minkowskian space-time is also the solution
to the whole system of the equations of motion.

Consider a small scalar field
$$
  X\sim\e,~~~~P\sim\e,~~~~\e\ll1.
$$
Then $T\sim\e^2$, and the first correction to $W_m$ and hence to the
metric is of the second order $\e^2$ too. Therefore for a small scalar
field effective equations of motion are linear spherical wave equations
in the background given by the Schwarzschild or Lorentz metric
(\ref{eshwme}). This is a good approximation because corrections to
the metric are of the second order. In fact, this can be clearly seen at
the very beginning because the energy-momentum tensor is quadratic in a
scalar field and it is the source for corrections to the metric.

Now we show how the effective action (\ref{effacs}) reduces to the action
of a free scalar field in the Minkowskian space-time or black hole
background. For a small scalar field
$$
  e^{2T}\approx 1+2T.
$$
Inserting this approximation into the effective Hamiltonian
(\ref{effacs}) for $A_m=0$ we get
\begin{equation*}
  H_{\rm eff}=-\int_0^\infty\!dr T
  =\frac12\int_0^\infty\!dr\int_r^\infty\!\frac{d\s}\s
  \left(\frac1{\s^2}P^2+\s^2\pl_1X^2\right).
\end{equation*}
Exchanging the order of integration one arrives at the usual Hamiltonian
for a spherically symmetric waves in Minkowskian space-time
\begin{equation}                                        \label{effacu}
  H_{\rm eff}
  =\frac12\int_0^\infty\!dr
  \left(\frac1{r^2}P^2+r^2\pl_1X^2\right).
\end{equation}
This effective action is clearly positive definite. Here we see the
role of the normalization unity in the integrand in (\ref{effacs}).
Similar calculations for a small scalar field in the Schwarzschild
background $A_m=-2M$ reduce the effective Hamiltonian (\ref{effacs}) to
\begin{equation}                                        \label{efascb}
  H_{\rm eff}
  =\frac12\int_0^\infty\!dr\left(1-\frac{2M}r\right)
  \left(\frac1{r^2}P^2+r^2\pl_1X^2\right).
\end{equation}
The obtained expression is positive definite outside the horizon of
a black hole $r>2M$.

At large distances $r\rightarrow\infty$ the effective action reduces
to spherical waves in a flat background too (wave zone). Assume that
all integrals are convergent at $r\rightarrow\infty$. Then
$T(t,r)\rightarrow0$ due to the upper limit of integration, and
$W_m\sim r$. Neglecting $A_m\ll r$ one gets the effective Hamiltonian
(\ref{effacu}) for spherical waves in flat Minkowskian space-time.
In the wave zone the metric (\ref{espemd}) reduces to a Lorentzian one.

At small distances $r\rightarrow0$ we have
$$
  W_m(t,r)\approx re^{2T_0},
$$
where $T_0=T(t,0)>-\infty$ is assumed to be finite. Then the effective
equations of motion and the metric become
\begin{align}                                           \label{eomxsr}
  \dot X&=\frac1{r^3}(A_m+re^{2T_0})P,
\\                                                      \label{eompsr}
  \dot P&=\pl_1\left[(A_m+re^{2T_0})r\pl_1X\right],
\\
  ds^2&=\frac1\kappa e^{2T_0}\left[(e^{2T_0}+{\textstyle\frac{A_m}r})dt^2
  -\frac1{e^{2T_0}+\frac{A_m}r}dr^2\right].
\end{align}
\subsection{The Fisher solution                        \label{stsosp}}
For spherically reduced gravity the effective equations of motion can
be analytically solved for the massless static scalar field $P=0$
\cite{Fisher48}. In this case Equation (\ref{efeqst}) becomes
\begin{equation}                                        \label{estsoe}
  Z''=\frac{Z'}{rZ^2},
\end{equation}
where
$$
  Z(r)=\frac1{r\pl_1X}.
$$
This equation has a general solution depending on two arbitrary constants
$k$ and $c$
\begin{equation}                                        \label{egesoy}
  r=\frac ky(y_1-y)^{\frac1{y_1^2+1}}(y-y_2)^{\frac1{y_2^2+1}},
\end{equation}
where
$$
  y=\frac1Z.
$$
Here
$$
  y_{1,2}=-\frac c2\pm\sqrt{\frac{c^2}4+1},
$$
are the roots of the quadratic equation
\begin{equation}                                        \label{erootz}
  y^2+cy-1=0
\end{equation}
which has real roots $y_1>0$ and $y_2<0$ for arbitrary values of $c=\const$.
The constant $k$ corresponds to a scaling symmetry and is insignificant.

In this way we solved the effective equations of motion for a scalar
field in the static case. To write the corresponding metric in elementary
functions we choose $y$ as a spacial coordinate instead of $r$. For
$r\in(0,\infty)$ the new coordinate varies within the interval
$y\in(0,y_1)$ but in reverse direction. Straightforward calculations
yield nonlocal quantities (\ref{etspgr}), (\ref{exdilw}) and a
scalar field
\begin{align}                                            \nonumber
  2T&=\frac1{y_1-y_2}\ln\frac{(1-\frac y{y_1})^{y_1}}{(1-\frac y{y_2})^{y_2}},
\\                                                       \nonumber
  W&=ky_1^{-\frac{y_1^2}{y_1^2+1}}(-y_2)^{-\frac{y_2^2}{y_2^2+1}}
  \left(\frac1y-\frac1{y_1}\right),
\\ \intertext{and the expression for a scalar field}     \label{escycs}
  X&=\pm\frac1{y_1-y_2}\ln\left|\frac{y-y_1}{y-y_2}\right|.
\end{align}
The scalar field is always singular at the origin of spherical
coordinates $r=0,$ or $y=y_1$. Substitution of this solution directly
into Equation (\ref{emzest}) yields $A_m=k/y_1$
(this is a necessary step because this equation was differentiated).

Dropping the gravitational constant and the insignificant total factor
depending on the roots $y_{1,2}$ one finally arrives at the static
solution
\begin{align}                                        \nonumber
  ds^2&=(y_1-y)^{\frac{y_1^2-1}{y_1^2+1}}
  (y-y_2)^{\frac{y_2^2-1}{y_2^2+1}}dt^2
  -\frac{k^2}{y^4}(y_1-y)^{-\frac{y_1^2-1}{y_1^2+1}}
  (y-y_2)^{-\frac{y_2^2-1}{y_2^2+1}}dy^2
\\                                                   \label{estmeu}
  &-\frac{k^2}{y^2}(y_1-y)^{\frac2{y_1^2+1}}(y-y_2)^{\frac2{y_2^2+1}}
  d\Om.
\end{align}
This expression together with (\ref{escycs}) solves the problem
and describes all static spherically symmetric solutions for
general relativity minimally coupled to a massless scalar field.
Up to a rescaling a general solution is parameterized by one
arbitrary constant $c$ defining the roots (\ref{erootz}).

To get physical interpretation of these solutions one has to calculate
the {\it four-dimensional} curvature components. Nonzero components of
Christoffel's symbols are
\begin{align*}
  \G_{ty}{}^t&=\G_{yt}{}^t=-\frac{y_1+y_2}{2(y_1-y)(y-y_2)},
\\
  \G_{tt}{}^y&=-\frac{y^4}{2k^2}(y_1-y)^{2\frac{y_1^2-1}{y_1^2+1}}
  (y-y_2)^{2\frac{y_2^2-1}{y_2^2+1}}\frac{y_1+y_2}{(y_1-y)(y-y_2)},
\\
  \G_{yy}{}^y&=-\frac2y+\frac{y_1+y_2}{2(y_1-y)(y-y_2)},
\\
  \G_{y\theta}{}^\theta&=\G_{\theta y}{}^\theta
  =\G_{y\f}{}^\f=\G_{\f y}{}^\f=
  -\frac1{y(y_1-y)(y-y_2)},
\\
  \G_{\theta\theta}{}^y&=y,
\\
  \G_{\theta\f}{}^\f&=\G_{\f\theta}{}^\f=\frac{\cos\theta}{\sin\theta},
\\
  \G_{\f\f}{}^y&=y\sin^2\theta,
\\
  \G_{\f\f}{}^\theta&=-\sin\theta\cos\theta.
\end{align*}
Afterwards we compute the nonzero curvature components
\begin{align*}
  R_{tyty}&=-\frac{y_1+y_2}y
  (y_1-y)^{-\frac{y_1^2+3}{y_1^2+1}}
  (y-y_2)^{-\frac{y_2^2+3}{y_2^2+1}}
\\
  R_{t\theta t\theta}&=\frac{y(y_1+y_2)}2
  (y_1-y)^{-\frac2{y_1^2+1}}
  (y-y_2)^{-\frac2{y_2^2+1}},
\\
  R_{t\f t\f}&=R_{t\theta t\theta}\sin^2\theta,
\\
  R_{y\theta y\theta}&=\frac{k^2(2y-y_1-y_2)}{2y^3}
  (y_1-y)^{-\frac{2y_1^2}{y_1^2+1}}
  (y-y_2)^{-\frac{2y_2^2}{y_2^2+1}},
\\
  R_{y\f y\f}&=R_{y\theta y\theta}\sin^2\theta,
\\
  R_{\theta\f\theta\f}&=-\frac{k^2(y-y_1-y_2)}y
  (y_1-y)^{-\frac{y_1^2-1}{y_1^2+1}}
  (y-y_2)^{-\frac{y_2^2-1}{y_2^2+1}}\sin^2\theta.
\end{align*}
The Ricci tensor has only one nonvanishing component though many
components of the full curvature tensor differ from zero
\begin{equation*}
  R_{yy}=-\frac2{(y_1-y)^2(y-y_2)^2}.
\end{equation*}
Now one can easily verify that the metric (\ref{estmeu}) together with
the scalar field (\ref{escycs}) satisfies Einstein equations
\begin{equation}                                        \label{eineqx}
  R_{ij}=-2\pl_i X\pl_j X,
\end{equation}
for all indices $i,j=0,1,2,3$. Note that the Ricci tensor is invariant
under rescaling of the metric. Therefore dropping a constant factor in
the solution for the metric (\ref{estmeu}) does not alter this equation.

The four-dimensional scalar curvature for static solutions has the form
\begin{equation}                                        \label{estscx}
  R=\frac{2y^4}{k^2}(y_1-y)^{-\frac{y_1^2+3}{y_1^2+1}}
  (y-y_2)^{-\frac{y_2^2+3}{y_2^2+1}}.
\end{equation}
At the origin of the spherical coordinate system $y=y_1$ the scalar
curvature is singular because the exponent is always negative
$$
  -\frac{y_1^2+3}{y_1^2+1}<0.
$$
Note that here the scalar field is also singular.

At infinity $y=0$ the scalar curvature tends to zero. The metric
(\ref{estmeu}) is degenerate at this point. Nevertheless one can show
that in the spherical coordinates $t,r,\theta,\f$ the metric is
asymptotically Lorentzian at infinity. It means that static solutions
(\ref{estmeu}) describe maximally extended space-times with naked
timelike singularity at the origin. The $t,r$ slices are represented
by the triangular Carter--Penrose diagram.

Static solution for $c=0$ has a particular simple form and may be
written in spherical coordinates explicitly. In this
case Equation (\ref{egesoy}) takes the form
$$
  \frac{Z-1}r=\frac r{Z+1}.
$$
It yields the solution for the scalar field and
nonlocal quantities
\begin{align}                                         \label{esoscf}
  X&=\pm\frac12\ln\frac{\sqrt{1+r^2}-1}{\sqrt{1+r^2}+1},
\\                                                    \nonumber
  2T&=\frac12\ln\frac{r^2}{1+r^2},
\\                                                    \nonumber
  W_m&=\sqrt{1+r^2}-1.
\end{align}
Now one easily writes the solution for a metric (\ref{espemd})
\begin{equation}                                        \label{estmes}
  ds^2=dt^2-\frac{r^2}{1+r^2}dr^2-r^2d\Om.
\end{equation}
This metric is degenerate at the origin and asymptotically flat at
infinity. A recently discovered family of nonstatic solutions
\cite{Robert89} intersects with static solutions by this representative.

The scalar curvature (\ref{estscx}) for $c=0$ becomes
\begin{equation}                                        \label{efdscc}
  R=\frac2{r^4}.
\end{equation}

Let us note that the two-dimensional part of the metric (\ref{estmes}) by
itself can be easily transformed to the Lorentz metric. Therefore
two-dimensional $t,r$ curvature components identically vanish, and there
is no singularity in the two-dimensional curvature. The singularity in
(\ref{efdscc}) comes from the mixed radial-angular and angular components
of the full four-dimensional curvature tensor.
\subsection{The Roberts solution                          \label{srobso}}
In this section we consider the Roberts solution \cite{Robert89}
providing a nontrivial solution to the effective equations of motion
(\ref{escspx}), (\ref{escspp}) depending both on time and space coordinates.
It is usually written in the form
\begin{align}                                           \label{erosox}
  X&=\frac12\ln\frac{\sqrt{a^2u^2+r^2}-au}{\sqrt{a^2u^2+r^2}+au}
    =\ln\left(\sqrt{1+a^2\frac{u^2}{r^2}}-a\frac ur\right),
\\                                                      \label{erosom}
  ds^2&=\left(1-\frac{2a^2u}{\sqrt{a^2u^2+r^2}}\right)du^2
  -\frac{2r}{\sqrt{a^2u^2+r^2}}drdu-r^2d\Om,
\end{align}
where $a=\const\ne0$ and $u$ is a light-cone coordinate. Without loss of
generality we consider positive $a>0$ because equations of motion
(\ref{eqmopi})--(\ref{eqmozx}) are invariant under the transformation
$X\rightarrow-X$. In the wave zone corresponding to the limit
$$
  r\rightarrow\infty,~~~~~~\frac ur\rightarrow0,
$$
the time-radial part of the metric (\ref{erosom}) becomes
\begin{equation}                                        \label{easmer}
  ds^2=\left(1-\frac{2a^2u}r\right)du^2-2drdu,
\end{equation}
where we retained the first correction to the Lorentz metric,
and the leading term for the scalar field (\ref{erosox}) takes the form
$$
  X\approx-a\frac ur=-a\frac{t+r}r.
$$
In this domain the scalar field indeed represents the free ingoing
spherical wave in accord with a general statement in Section
\ref{smscfi}. We see that the Roberts solution describes asymptotically
Minkowskian space-time. Note that asymptotic behavior of the Roberts
metric (\ref{easmer}) differs from the Schwarzschild metric,
which in the Eddington--Finkelstein coordinates has the form
$$
  ds^2=\left(1-\frac{2M}r\right)du^2-2drdu.
$$

The Roberts solution takes a particular simple form in the light-cone
coordinates which are useful for the analysis of the global structure
of the space-time. Introducing the second light-cone coordinate $v$
related to $r$ and $u$ by the equation
\begin{equation}                                        \label{ereruv}
  r^2=\frac14(u-v)^2-a^2u^2,
\end{equation}
the solution takes the form
\begin{align}                                           \label{exnucr}
  X&=\frac12\ln\frac{(1-2a)u-v}{(1+2a)u-v}, & u&>v,
\\                                                      \label{exnusr}
  X&=\frac12\ln\frac{(1+2a)u-v}{(1-2a)u-v}, & u&<v,
\\                                                      \label{eromuv}
  ds^2&=dudv-r^2d\Om. &&
\end{align}
By definition the right hand side of Equation (\ref{ereruv}) must be
nonnegative. This restricts the possible range of lightlike coordinates
\begin{equation}                                        \label{eralco}
  [(1+2a)u-v][(1-2a)u-v]\ge0.
\end{equation}
For definiteness we consider the domain where both multipliers are
nonnegative. In this domain $u$ must be greater than $v$.

The lines $r=\const$ on the $u,v$ plane are deformed "hyperbolas"
given by Equation (\ref{ereruv}).

The global structure of the space-time depends on the curvature
singularities. Using Einstein equations for a massless scalar field
(\ref{eineqx}) one easily obtains expression for the four-dimensional
scalar curvature
$$
  R^{(4)}=-2\pl X^2,
$$
which for the Roberts solution becomes
\begin{equation}                                        \label{escros}
  R^{(4)}=\frac{2a^2uv}{r^4}.
\end{equation}
The scalar curvature is always singular at $r=0$ which on the $u,v$ plane
corresponds to two crossing straight lines
$$
  u=(1\pm2a)v.
$$
At space infinity $r\rightarrow\infty$ the scalar curvature tends to zero.

To check that the Roberts solution really satisfies the effective
equations of motion obtained in the previous sections one has to
transform the metric (\ref{erosom}) to the diagonal form (\ref{espemd}).
To this end the function $u=u(t,r)$ must satisfy the differential
equation
\begin{equation}                                        \label{equitr}
  \frac{\pl u}{\pl r}=\frac r{\sqrt{a^2u^2+r^2}-2a^2u}.
\end{equation}
Solution of this equation depends on the value of constant $a$.
There are three cases $0<a<1/2$, $a=1/2$, and $a>1/2$.
Omitting all calculations we summarize the result.
\subsubsection{The case $0<a<1/2$}
The transformation from $u,r$ to $t,r$ coordinates is given
by a solution of Equation (\ref{equitr})
\begin{equation}                                        \label{efifun}
  \Phi_1=\left|\sqrt{a^2u^2+r^2}-z_1u\right|^{z_1}
  -C(t)\left|\sqrt{a^2u^2+r^2}-z_2u\right|^{z_2}=0,
\end{equation}
where
$$
  z_{1,2}=\frac{1\pm\sqrt{1-4a^2}}2>0
$$
are the roots of the quadratic equation
$$
  z^2-z+a^2=0,
$$
and $C(t)$ is an arbitrary function of time $t$ with $\dot C\ne0$. The
latter appears because the diagonal gauge is defined up to an arbitrary
transformation of time coordinate. Straightforward calculations yield
the nontrivial metric components
\begin{align}                                           \label{egzzfc}
  g_{00}&=\left(\frac{\dot C}C\right)^2
  \frac{(2a^2u^2+r^2-u\sqrt{a^2u^2+r^2})^2}
  {(1-4a^2)\sqrt{a^2u^2+r^2}(\sqrt{a^2u^2+r^2}-2a^2u)},
\\                                                      \label{egoofc}
  g_{11}&=-\frac{r^2}{\sqrt{a^2u^2+r^2}(\sqrt{a^2u^2+r^2}-2a^2u)}.
\end{align}
Comparing these expressions with the metric (\ref{espemd}) one finds
\begin{align}                                           \label{eawfca}
  e^{2T}&=\frac{\dot C}C
  \frac{r(2a^2u^2+r^2-u\sqrt{a^2u^2+r^2})}
  {\sqrt{1-4a^2}\sqrt{a^2u^2+r^2}(\sqrt{a^2u^2+r^2}-2a^2u)}.
\\                                                      \label{eetfca}
  A_m+W_m&=\frac{\dot C}C
  \frac{2a^2u^2+r^2-u\sqrt{a^2u^2+r^2}}{\sqrt{1-4a^2}},
\end{align}
Using the equation of motion (\ref{escspx}) one easily finds the momentum
for a scalar field
\begin{equation}                                        \label{emoxfc}
  P=\frac{ar^3}{\sqrt{a^2u^2+r^2}(\sqrt{a^2u^2+r^2}-2a^2u)}.
\end{equation}
Straightforward calculations show that expressions (\ref{eawfca}) and
(\ref{eetfca}) are indeed related to the scalar field (\ref{erosox})
and its conjugate momentum (\ref{emoxfc}) through the integrals
(\ref{etspgr}) and (\ref{exdilw}) if and only if
\begin{equation}                                        \label{exprco}
  C(t)=t^{z_1-z_2}.
\end{equation}
This happened because in obtaining the effective action we have already
fixed the time coordinate. The last equation of motion (\ref{escspp}) is
checked by direct calculations. This proves that the Roberts solution
is the exact solution of the effective equations of motion.

Using Equation (\ref{exprco}) the transition function $\Phi_1$, the
temporal metric component, and nonlocal quantities may be written in
the form
\begin{align}                                           \label{exppho}
  \Phi_1&=\left|\frac{\sqrt{a^2u^2+r^2}-z_1u}t\right|^{z_1}
  -\left|\frac{\sqrt{a^2u^2+r^2}-z_2u}t\right|^{z_2}=0,
\\                                                      \label{egzffc}
  g_{00}&=\frac{(2a^2u^2+r^2-u\sqrt{a^2u^2+r^2})^2}
  {t^2\sqrt{a^2u^2+r^2}(\sqrt{a^2u^2+r^2}-2a^2u)},
\\                                                      \label{eaffca}
  e^{2T}&=\frac{r(2a^2u^2+r^2-u\sqrt{a^2u^2+r^2})}
  {t\sqrt{a^2u^2+r^2}(\sqrt{a^2u^2+r^2}-2a^2u)}.
\\                                                      \label{eeffca}
  A_m+W_m&=\frac1t(2a^2u^2+r^2-u\sqrt{a^2u^2+r^2}).
\end{align}
\subsubsection{The case $a=1/2$}
In the second case the transformation to the diagonal gauge is given by
the equation
\begin{equation}                                        \label{esetrt}
  \Phi_2=\frac12\sqrt{u^2+4r^2}-\frac12u
  -t\exp\left(\frac u{\sqrt{u^2+4r^2}-u}\right)=0,
\end{equation}
where we fixed an arbitrary function of time by the same procedure
as in the previous section. Afterwards we obtain expressions for the
metric components, nonlocal quantities, and momenta
\begin{align}                                           \label{emeszz}
  g_{00}&=\frac{(\sqrt{u^2+4r^2}-u)^3}
  {4t^2\sqrt{u^2+4r^2}},
\\                                                      \label{emesoo}
  g_{11}&=-\frac{\sqrt{u^2+4r^2}+u}{\sqrt{u^2+4r^2}},
\\                                                      \label{etesca}
  e^{2T}&=\frac{r(\sqrt{u^2+4r^2}-u)}{t\sqrt{u^2+4r^2}},
\\                                                      \label{ewasca}
  A_m+W_m&=\frac{(\sqrt{u^2+4r^2}-u)^2}{4t},
\\
  P&=\frac{2r^3}{\sqrt{u^2+4r^2}(\sqrt{u^2+4r^2}-u)}.
\end{align}
These expressions follow from Equations (\ref{egzffc})--(\ref{eeffca}),
(\ref{egoofc}), and (\ref{emoxfc}) for $a=1/2$.
\subsubsection{The case $a>1/2$}
In the third case the transformation to the diagonal gauge is defined
by the equation
\begin{equation}                                        \label{ethctr}
  \Phi_3=\sqrt{2a^2u^2+r^2-u\sqrt{a^2u^2+r^2}}
  -t\exp\left(\frac1b\arctg\frac{bu}{2\sqrt{a^2u^2+r^2}-u}\right)=0,
\end{equation}
where we introduced a shorthand notation
$$
  b=\sqrt{4a^2-1}=\const.
$$
Explicit expressions for the metric components, nonlocal quantities,
and momentum are the same as before (\ref{egoofc}), (\ref{emoxfc}),
(\ref{egzffc})--(\ref{eeffca}). Note that $\Phi_3\rightarrow\Phi_2$
for $a\rightarrow1/2$.

We see that in all three cases expressions for geometric quantities
are the same formally, the only difference being the transition
functions (\ref{exppho}), (\ref{esetrt}), and (\ref{ethctr}). The
solution is obviously continuously self-similar.

The Roberts solution is in striking conflict with the intuition
gained from the Schwarzschild solution. For example, the
line defined by the equation
$$
  \frac1{g_{11}}=0~~~~\Leftrightarrow~~~~v=(1-4a^2)u,
$$
which is often assumed to define a horizon is not lightlike for
$a\ne1/2$. At the singularity $r=0$ the temporal component of the
metric is a nonzero constant for $a\ne1/2$ and equals zero for
$a=1/2$. The radial component of the metric $g_{11}=0$ at $r=0$
for $a\ne0$.

The Roberts solution does not describe a formation of a black hole
because a true singularity at the origin $r=0$ exists for all times.
On the $u,v$ plane it is timelike in the past $u<0$ and timelike,
lightlike, or spacelike in the future $u>0$ for $0<a<1/2$, $a=1/2$,
or $a>1/2$, respectively. In the last two cases the type of the
singularity changes during the evolution.
\section{Conclusion}
In the present paper we obtain the effective action for physical
degrees of freedom in a general two-dimensional gravity model coupled
to a scalar field by a step by step explicit solution of the equations
of motion and elimination of all unphysical geometric variables.
The resulting effective action contains only a scalar field and its
conjugate momentum and is written in an abstract Minkowskian space-time.
All geometric variables defining the geometry of the real space-time
are given by simple formulas in terms of a scalar field. The whole
problem of the solution of the equations of motion is reduced to the
following steps. First, one has to find an exact or approximate
solution to the effective equations of motion for a scalar field.
Second, the metric and other geometrical quantities should be
computed. Third, one should analyze the physical properties of the
space-time. We checked that the Fisher and Roberts exact solutions are
indeed the solutions to the effective equations of motion.

The resulting effective action is nonlocal and complicated, and it is
not clear whether it simplifies the solution of the equations of motion
or not. In any case its existence guarantees the existence of the
conserved quantity which for solutions with the Schwarzschild
asymptotic may be identified with the total mass of the space-time.

The effective action was obtained locally by the solution of the
equations of motion, and its existence does not depend on the
global structure of the space-time. It appears as a boundary term,
and the reason for its appearance lies in the constraints. We showed
that in general a solution of the constraints does not have a compact
support, and to produce the equations of motion without boundary
conditions on the fields one must add the boundary term to the action.
This boundary term does not depend on the action we started with and is
defined entirely by the constraints.

For closed universes the assumption of smoothness for all the fields is
in contradiction with a solution of the constraints: There are no such
solutions. Therefore this assumption must be weakened. We assume that only
physical degrees of freedom are smooth functions in a closed universe.
The unphysical degrees of freedom are nonsmooth functions in this case
as a solution of the constraints, and the corresponding boundary term
must be added to the action. This boundary term is responsible for the
effective action which correctly reproduces the effective equations of
motion, and its numerical value may be used for the definition of the
total energy even for closed universes.

The importance of boundary terms for the definition of the total energy
for asymptotically flat space-times was noted long ago
\cite{ArDeMi62,RegTei74} (for review see \cite{Faddee82,GitTyu90}).
Roughly speaking, the conclusion was based on two important observations.
First, one of the boundary terms which is dropped in obtaining the
equations of motion equals the mass $M$ for the Schwarzschild
solution which is assumed to yield a correct asymptotic for any compact
distribution of matter. Second, it reproduces the correct equations
of motion for physical degrees of freedom in the linear approximation
in which the constraints in general relativity may be solved. In the
present paper we obtain the same boundary term in full generality
without using any asymptotic or approximation. This derivation shows
two interesting features: The boundary term is nontrivial even for
closed universes, and on the constraints it is given as an integral
of some density over the whole space-time as usual. This is because
the boundary term contains values of unphysical degrees of freedom
on the boundary which are in their turn given by the integral of
physical degrees of freedom over the space as a solution to the
constraints.

We considered a wide class of two-dimensional gravity models arbitrary
coupled to a scalar field which includes the spherically reduced general
relativity. This suggests that the situation with the constraints is
quite general: The constraints in a gravity model do not admit solutions
with compact support, and the boundary term must be added to the action.
This boundary term results in the effective Hamiltonian for physical
degrees of freedom and should be used for the definition of the total
energy of the space-time. In the present paper we have found it by
explicit solution of the constraints. The problem remains of how to
obtain it without a
solution of the constraints. This is important, for example,
for the path integral quantization. Indeed, the measure in the path
integral \cite{FraVil77} in our case contains $6$ $\dl$-functions
corresponding to the constraints (\ref{elcmze}), (\ref{ezwmze}),
(\ref{esecon}), and (\ref{esectw}) and $6$ $\dl$-functions with gauge
conditions (\ref{egacon}), (\ref{etizws}), and (\ref{etclos}) which
together form a set of $12$ second class constraints. Solution of
the constraints and gauge conditions results in a zero effective
Hamiltonian if the boundary term is not added from the beginning.
We know that it must be added because otherwise the correct equations
of motion for the physical degrees of freedom are lost, and the
problem is what to add if we are not able to solve the constraints
explicitly which is the case for more general gravity models.
This problem has a direct influence for the correct covariant
perturbation theory.

Many thanks to D.~Grumiller, W.~Kummer, I.~V.~Tyutin, and I.~V.~Volovich
for fruitful discussions. The financial support of the Russian Foundation
for Basic Research is greatly acknowledged, grants RFBR 96-15-96131 and
99-01-00866. The author thanks the Austrian Academy of
Science and Technical University of Vienna where this work began for
the hospitality.
\section*{Appendix. Dimensions of the fields           \label{sdifie}}
The analysis of the equations of motion is cumbersome. Therefore we
attribute dimensions to every field and coupling constant to be able
to check dimensions of expressions at each step of the calculations.
By definition, coordinates have dimension of length $[x^\al]=l$,
the zweibein components and the action are dimensionless
$$
  [e_\al{}^a]=1,~~~~[S]=1,
$$
and the Lorentz connection has the same dimension as the partial derivative
$\pl_\al$
$$
  [\om_\al]=[\pl_\al]=l^{-1}.
$$
This means that torsion components and the scalar curvature have dimensions
$$
  [T^{*a}]=l^{-1},~~~~[R]=l^{-2}.
$$
We assume also that a scalar field has the same dimension as in
four-dimensional space-time
\begin{equation}                                        \label{ediscf}
  [X]=l^{-1},
\end{equation}
because in section \ref{sredgr} we analyze spherically reduced gravity.
Afterwards dimensions of all other variables are uniquely defined.
A Lagrangian in two dimensions has dimension $[L]=l^{-2}$. Hence
\begin{equation}                                        \label{edippi}
  [\pi]=1,~~~~[p_a]=l^{-1}.
\end{equation}
Arbitrary functions $U(\pi)$ and $V(\pi)$ entering the Lagrangian
(\ref{egrato}) have dimensions
\begin{equation}                                        \label{edimuv}
  [U]=1,~~~~[V]=l^{-2}.
\end{equation}
So one could introduce a dimensionfull coupling constant in front of
$V(\pi)$. The dimension of the mass of a scalar field (\ref{elagst})
is the same in any dimensional space-time
$$
  [m]=l^{-1}.
$$
The dimension of the coupling $\rho$ is defined by the dimension of a
scalar field (\ref{ediscf}) and dimensionlessness of the action
$$
  [\rho]=l^2.
$$
The four-dimensional gravitational coupling constant $\kappa$ has dimension
$$
  [\kappa]=l^{-2}.
$$

For minimally coupled scalars in two dimensions one would have
$$
  [X]=1,~~~~[\rho]=1.
$$
These dimensions are suitable for the analysis of a bosonic string with
dynamical geometry when the gravity Lagrangian (\ref{egrato}) is added
to the string model \cite{KatVol86}. We do not consider this model
in the present paper.

\end{document}